\documentclass[final,3p,times,onecolumn]{elsarticle}
\usepackage{graphicx}
\usepackage{amssymb,amsmath,color}
\usepackage{subfigure}

\usepackage{epstopdf}
\usepackage[ruled]{algorithm2e}

\newcommand{\trademark}[1]{#1\textsuperscript{\textregistered}}

\usepackage{array,booktabs}
\newcolumntype{L}{@{}>{\kern\tabcolsep}l<{\kern\tabcolsep}}
\usepackage{colortbl}
\usepackage{hyperref}
\usepackage{color}
\usepackage{xcolor}
\definecolor{dark-red}{rgb}{0.4,0.15,0.15}
\definecolor{dark-blue}{rgb}{0.15,0.15,0.4}
\definecolor{medium-blue}{rgb}{0,0,0.5}
\definecolor{bananamania}{rgb}{0.98, 0.91, 0.71}
\definecolor{}{rgb}{1.0, 0.97, 0.91}

\hypersetup{
  colorlinks   = true, %Colours links instead of ugly boxes
  urlcolor     = blue, %Colour for external hyperlinks
  linkcolor    = blue, %Colour of internal links
  citecolor   = red %Colour of citations
}

\definecolor{darkred}{rgb}{0.4,0.15,0.15}
\definecolor{darkblue}{rgb}{0.15,0.15,0.4}
\definecolor{medium-blue}{rgb}{0,0,0.5}
\definecolor{darkgreen}{rgb}{0.0, 0.5, 0.0}
\definecolor{aqua}{rgb}{0.0, 1.0, 1.0}

\journal{Journal of Non-Newtonian Fluid Mechanics}

\date{}                                           

\journal{Journal of Non-Newtonian Fluid Mechanics}

\begin{document}

\begin{frontmatter}

\title{Microscopic flows of suspensions of the green non-motile \textit{Chlorella} micro-alga at various volume fractions: applications to intensified photo bio reactors}

\author[LTN]{Antoine Souli\`{e}s}
%\ead{Antoine.Soulies@etu-univ-nantes.fr}

\author[GEPEA]{J\'{e}r\'{e}my Pruvost}

\author{Cathy Castelain,\fnref{LTN}}
%\ead{Cathy.Castelain@univ-nantes.fr}

%\author[LTN]{Teodor Burghelea }

%\ead{Teodor.Burghelea@univ-nantes.fr}

%\address[LTN]{Universit\'{e} de Nantes, CNRS, Laboratoire de Thermocin\'{e}tique, UMR 6607, La Chantrerie,  Rue Christian Pauc, B.P. 50609, F-44306 Nantes Cedex 3, France}
%\address[]{}

\author{Teodor Burghelea \corref{cor1} \fnref{LTN}}
\cortext[cor1]{Corresponding Author: Teodor Burghelea  \ead{Teodor.Burghelea@univ-nantes.fr}, Tel.: + (33) - (0)2 40 68 31 85; Fax: + (33) - (0)2 40 68 31 41}
\fntext[LTN]{Universit\'{e} de Nantes, CNRS, Laboratoire de Thermocin\'{e}tique, UMR 6607, La Chantrerie,  Rue Christian Pauc, B.P. 50609, F-44306 Nantes Cedex 3, France}
\fntext[GEPEA]{Universit\'{e} de Nantes, CNRS, Genie des Proc\'{e}d\'{e}s - Environnement  Agroalimentaire, UMR 6144, 37 boulevard de l'{U}niversit\'{e}, BP 406, 44602 Saint-Nazaire Cedex, France}

\begin{abstract}

An experimental study of flows of the green non-motile \textit{Chlorella} micro-alga in a plane microchannel is presented. Depending on the value of the cell volume fraction, three distinct flow regimes are observed. For low values of the cell volume fraction a Newtonian flow regime characterised by a Poiseuille like flow field, absence of wall slip and hydrodynamic reversibility of the flow states is observed. For intermediate values of the cell volume fraction, the flow profiles are consistent with a Poiseuille flow of a shear thinning fluid in the presence of slip at the channel's wall. For even larger cell volume fractions, a yield stress like behaviour manifested through the presence of a central solid plug is observed. Except for the Newtonian flow regime, a strong hydrodynamic irreversibility of the flow and wall slip are found. The calculation of the wall shear rate and wall stress based on the measured flow fields allows one to identify the mechanisms of wall slip observed in the shear thinning and yield stress regimes.

\end{abstract}


\end{frontmatter}

\clearpage
\tableofcontents
\hypersetup{linkcolor=red}
\clearpage
\listoffigures
\hypersetup{linkcolor=red}
\clearpage
\section{Introduction} \label{sec:introduction}

Micro-algae are a large and biologically diverse group of aquatic micro-organisms with a relatively simple uni-cellular structure that can be found in various environments ranging from freshwater for some species to sea water for others. Most micro-algae species are photoautotrophic which means they convert solar energy into chemical forms through photosynthesis. 

During the past several decades the micro-algae have received a considerable amount of interest due to their potential use in several key industries 
related to food, cosmetics and \textit{"green"} energy. From an economical perspective their most appealing application is undoubtedly that as a potential feedstock for the biofuel production. This is because micro-algae may generally produce polysaccharides (sugars) and triacylglycerides (fats) which are the raw materials for producing bioethanol and biodiesel fuels. 
Whether the micro-algae can make it as a viable \textit{"green"} energy alternative in the near future is an intricate matter of both economic and energetic efficiency, \cite{Slade201329}. 

Among a variety of species of micro-algae, \textit{Chlorella Vulgaris (CCAP 211-19)} is a good candidate for the large scale production of biofuels and, for this reason, it has been intensively studied, \cite{chlorella1,chlorella2,chlorella3,chlorella4}. It is a non-motile alga (it has no flagellum) with a average diameter of roughly $4~\mu m$. Its membrane is quite rigid due to the presence of chitin-like glycan in the cell wall, \cite{membrane}. Thus, the \textit{Chlorella} micro-alga can sustain relatively high hydrodynamic stresses without any cellular damage, \cite{stresschlorella}.

There currently exist two main technologies for the cultivation of the micro-algae: \textit{raceway pond} systems and \textit{photobioreactors} (PBRs). A typical raceway pond is a closed  and shallow (with depth varying from $0.2 ~m$ to $0.5~m$) loop, open to the air. The large open surface allows an optimal exposure to light but it may also favour the formation of superficial films. In PBRs the culture medium is enclosed in a container with optically transparent walls and the micro-algal suspension is circulated from a central reservoir. Thus, the PBR systems allow for a better control culture environment but, in turn,  are getting more expensive than the traditional raceway pond systems. In addition to that, the energy consumption may also be higher than in the case of the raceway ponds,  \cite{Chisti2007294,Slade201329,Pulz,Schenk}.

However, due to a high degree of control of the culture conditions, the intensified PBR technology allows a higher productivity. With an appropriate engineering and operating protocol, a high cell density culture (biomass concentration) can be obtained. Hydrodynamics is one of the key aspects when working in high cell density culture. High cell density cultures are indeed obtained in systems having a very high ratio between the illuminated surface and the culture volume, which translates into a shallow depth, typically smaller than $0.01 m$,  \cite{cornet1,provost,doucha}. The high degree of geometric confinement combined with a high biomass concentration typically leads to a decrease of the mixing performances which can have several negative impacts on the process: a decrease of the efficiency of the mass transfer (mixing), an increase in the risk of biofilm formation and an overall less efficient light conversion in the systems due to smaller displacement of flowing cells along the light gradient in the culture volume. In this context, optimising the hydrodynamic conditions is of paramount practical importance.

A comprehensive review of the progress in the theoretical/numerical work dealing with the optimisation of the hydrodynamic conditions in intensified PBRs has been recently presented in Ref. \cite{Bitog2011131}. As most flows in intensified PBRs are characterised by a large Reynolds numbers ($Re$), a first difficulty encountered in the CFD studies of flows in PBRs is related to the proper choice of a turbulence model. Many authors chose the $k-\epsilon$ model which is implemented in commercially available CFD software packages such as FLUENT or Comsol Multiphysics, \cite{Pruvost20064476,cfd1}. 

The numerical studies of flows of micro-algae suspensions in PBRs may be challenging even at low $Re$ (\textit{i.e.} no inertial nonlinearity in the momentum equation), however. Whereas a dilute micro-algae  suspension may be well approximated by a Newtonian fluid characterised by a linear stress-rate of strain relationship, semi-dilute and concentrated suspensions which are relevant to many industrial processes \cite{highdensity} are characterised by a strongly nonlinear constitutive relationship.

\begin{figure}
\begin{center}
\centering
\includegraphics[width=0.5\textwidth, angle=90]{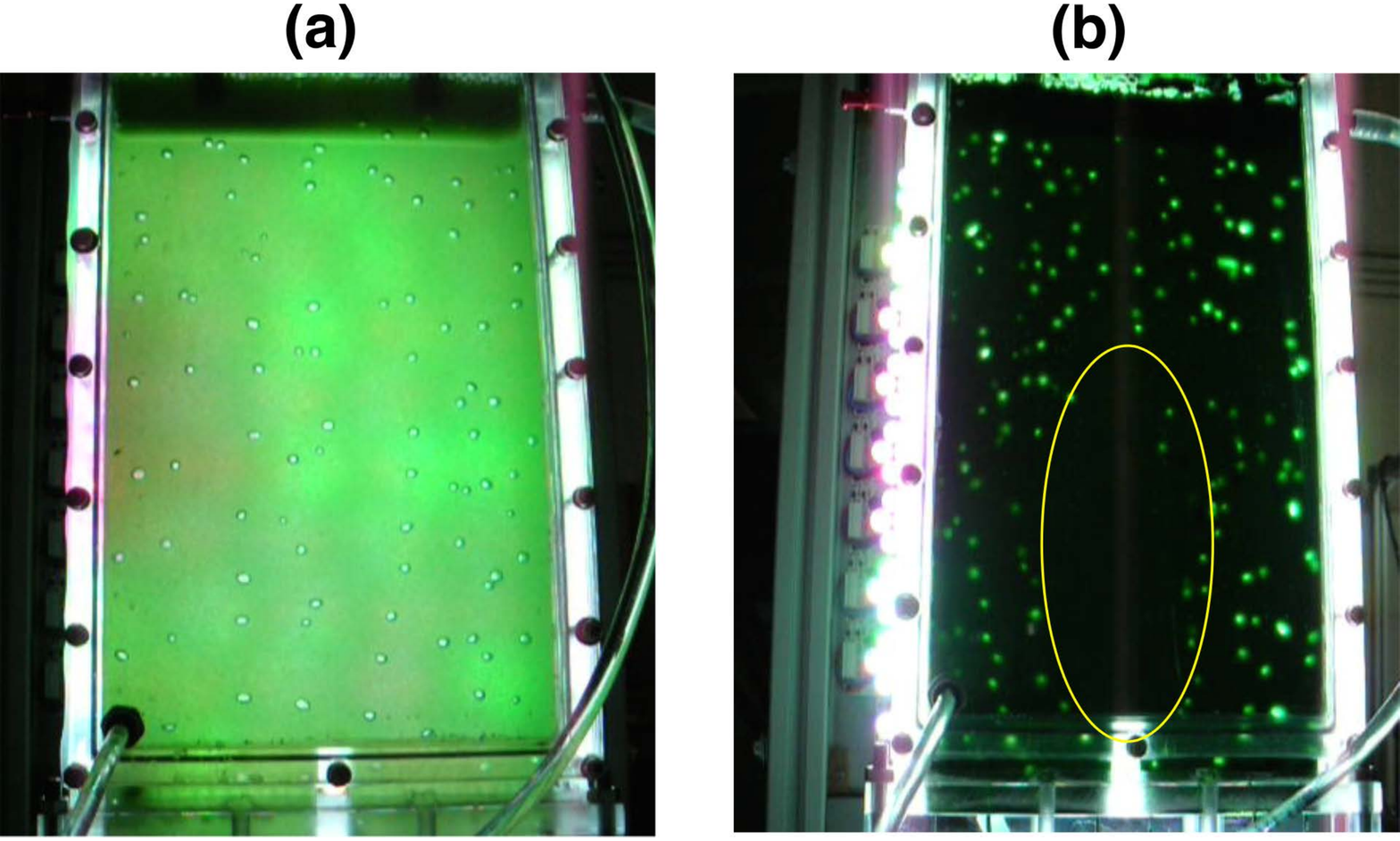}
\caption{\label{fig:bubbles} Images of gas bubbles raising within a flat panel PBR filled with a suspension of \textit{Chlorella} microalaga at two different concentrations: \textbf{(a)} $C_x \approx 0.5 g/l$ \textbf{(b)} $C_x \approx 5 g/l$. The closed curve in panel (b) highlights a dead flow zone. The animated versions of the pictures are available online as a supplemental material.  }
\end{center}
\end{figure}

To better understand the impact of the rheological properties of the micro-algae suspensions on their flow behaviour we illustrate in Fig. \ref{fig:bubbles} two photographs of a flat panel PBR containing a dilute (left panel) and a concentrated (right panel) suspension of the \textit{Chlorella} micro-alga being stirred by injection of gas bubbles from the bottom.

The dimensions (height, width, depth) of the PBR illustrated in Fig. \ref{fig:bubbles} are $25~cm~X~15~cm~X~7~mm$. The PBRs are illuminated from behind by a LED panel. The animated version of each snapshot is available online as a supplemental material. 
The stirring flow is generated by controlled injection of gas bubbles through orifices machined within the bottom plate of the container at equidistant locations. Due to their large buoyancy, the evenly injected gas bubbles raise and create local mixing which allows an optimal exposure of the micro-algae to the illuminated regions of the PBR located in the proximity of the optically transparent walls.
In the case of the dilute suspension ($C_x \approx 0.5 g/l$), the gas bubbles raise at locations which are more or less evenly distributed along the horizontal direction although their trajectories are not perfectly linear but show some oscillatory (swirling) behaviour (see the supplemental material).

   A fundamentally different motion of the gas bubbles is observed in the case of the concentrated solution ($C_x \approx 5 g/l$) in the form of a dead flow zone located around the central part of the PBR. As the gas bubbles are uniformly injected through the bottom plate of the PBR through equidistant orifices, the sole interpretation of this observation is that the rheological behaviour of the suspension is spatially non uniform (along the horizontal direction) and strongly non Newtonian (with a larger apparent viscosity) around the central part of the PBR. This very simple experimental observation performed in conditions which are relevant to the intensified PBR technology indicates that in a range of large micro-algae concentrations (which is the range of practical interest) both the rheology and the hydrodynamics significantly depart from the Newtonian behaviour.  

To our best knowledge, there exist only few previous studies of the rheological properties of suspensions of \textit{Chlorella} micro-alga. The rheological study by Wu and his coworker indicates that as the volume fraction is increased the rheological behaviour of suspensions of \textit{Chlorella pyrenoidosa} micro-alga changes from Newtonian like to non-Newtonian, \cite{wu}.   A more recent study of the rheological behaviour of suspensions of \textit{Chlorella Vulgaris} with volume fractions spanning a wide range was reported in Ref. \cite{rheochlorella}. 

In the context of the intensified PBR's, however, the relevance of rheometric flows is somewhat hard to define. The present study concerns with a systematic characterisation of flows of suspensions of \textit{Chlorella Vulgaris} in a plane micro-channel. As the actual tendency is to decrease the characteristic size of intensified PBR's down to a millimetre scale, we believe that a microscopic flow investigation would bring valuable insights that could contribute to an optimised designed of the PBR's.
The central aim of this study is to provide a systematic description of the microscopic flow behaviour in a wide range of cell volume fractions and correlate the results with the rheological properties of the suspensions. The paper is organised as follows.

The experimental setup, the experimental methods and the preparation of the micro-algae suspensions are detailed in Sec.  \ref{sec:experimental}. The experimental results are presented in Sec. \ref{sec:results}. A brief reminder of the rheological properties of \textit{Chlorella} suspensions at various volume fractions is presented in Sec. \ref{subsec:rheo}. A systematic description of the microscopic flows is presented in Sec. \ref{subsec:micro}.
The paper closes with a brief discussion of the main findings and their possible impact on the design of hydrodynamic optimisation of the intensified PBR's, Sec. \ref{sec:conclusions}.  

\section{Experimental setup and methods} \label{sec:experimental}

\subsection{Microscopic flow control and investigation} \label{subsec:microflow}

\begin{figure}
\centering
\includegraphics[width=0.6\textwidth]{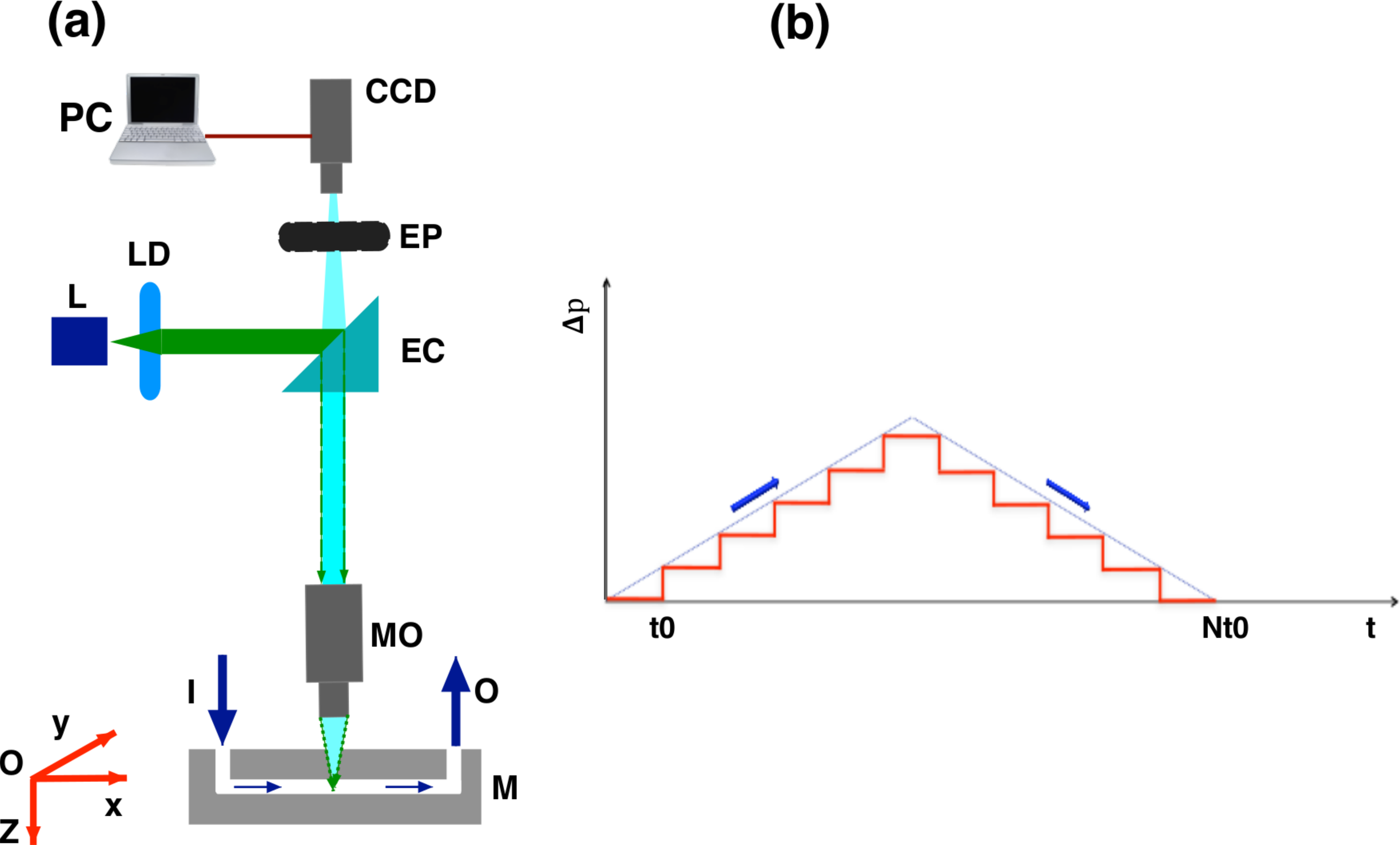}
\caption{\label{fig:setup} \textbf{(a) } Schematic representation of the experimental setup: \textbf{M} - micro channel, \textbf{I} - micro-channel inlet, \textbf{O} - micro-channel outlet, \textbf{MO} - microscope objective, \textbf{EP} - eye piece, \textbf{EC} - epifluorescent cube, \textbf{L} - light emitting diode (LED) panel,  \textbf{LD} - light diffuser, \textbf{CCD} - digital camera, \textbf{PC} - personal computer \textbf{(b)} Schematic representation of the controlled pressure ramp. $t_0$ is the characteristic forcing time (the averaging time per pressure value, see the explanation in the text) and $N$ is the total number of steps of the  ramp.}
\end{figure}

The experimental setup is schematically illustrated in Fig. \ref{fig:setup}(a). Our experiments have been conducted in a  borosilicate glass made micro-channel (from Micronit, Holland) with a rectangular cross section. The width of the micro channel is $W= 150~\mu m$ its depth is $H= 50~\mu m$ and its length is $L= 4~cm$. The corresponding hydraulic diameter is $D_h = 2 W H (W+H)^{-1} = 75~\mu m$. Based on the hydraulic diameter one can estimate the Reynolds number as $Re = \rho U D_h/ \eta$ where $U$ denotes the velocity scale and $\eta$ the viscosity scale. All the experiments discussed herein have been conducted at very low Reynolds numbers, $Re< 2.2 \cdot10^{-2}$. The suspensions feeding the micro-channel and collected at the outlet were held in plastic containers rigidly mounted on vertical rails. Each experimental campaign lasted less than $30~min$ so no significant evaporation took place at the level of the free surface within the inlet fluid container.
 
The microscopic flows were driven at a constant pressure drop $\Delta p$ controlled by controlling the elevation difference between the inlet and outlet fluid containers. To mimic the rheological procedure employed during our previous study reported in Ref. \cite{rheochlorella},  we  have performed controlled pressure drop linear ramps as represented schematically in Fig. \ref{fig:setup}(b). To modify the pressure drop, the elevation difference was modified by gently adding precise volumes of suspension to either the inlet or the outlet fluid container.
Corresponding to each step of the ramp, the pressure drop has been kept constant for a finite time interval $t_0$.  After reaching a maximum value the pressure drop is gradually decreased through the same intermediate steps. This allows one to probe the hydrodynamic reversibility of the flow and detect the possible emergence of hysteresis effect related to the competition between structuring and de-structuring of the micro-algal suspension. By choosing $t_0 = 6~s$ for all the experiments reported herein, we have deliberately placed ourselves in an unsteady flow condition. The reason behind this choice may be easily understood bearing in mind that the actual flows within the intensified PBR's are genuinely unsteady in the sense that the hydrodynamic stresses are locally exerted onto the volume elements of the suspension during times shorter than the characteristic times needed to reach a steady state deformation.  

The micro channel is mounted on an up-right microscope (Zeiss Axioscope $A_1$) as schematically illustrated in Fig.
\ref{fig:setup}(a). The microscope is equipped with a $20X$ magnification objective (Zeiss EC Plan Neofluar) and a depth of
field of $5.8~\mu m$.

Two distinct illumination schemes have been employed through our micro-fluidic experiments. To visualise the spatial distribution and the possible aggregation of individual \textit{Chlorella} cells during flow, a traditional bright field illumination has been used. For a quantitative  investigation of velocity fields, the micro flows have been illuminated in an epifluorescent mode. The excitation light is emitted by a LED panel L through a light diffuser, Fig. \ref{fig:setup}(a). The fluorescent emission is directed via an epifluorescent cube to a digital CCD camera (Chameleon from Point Grey). The flows were seeded with a minute amount ($\approx 100 ppm$) of fluorescent microspheres (Fluoresbrite multifluorescent from Polisciences) $0.9~\mu m$ in diameter. A significant difficulty in imaging concentrated suspensions of micro-algae is related to the light absorption. The attenuation of a monochromatic light beam traversing the suspension along the $z$ direction follows $I(z)=I_0exp(-z/z_a)$. The characteristic attenuation length $z_a$ depends on both the wave length of the incoming light $\lambda$ and the biomass concentration $C_x$, $z_a  =\left( E_a C_x\right )^{-1}$ where $E_a$ is the mass absorption coefficient which, for the \textit{Chlorella} micro-alga and $\lambda \approx 488  nm$ is $E_a \approx 300~m^2~kg^{-1}$.
For the largest biomass concentration we have used, $C_x = 143.7 g/l$, one can estimate $x_a \approx 23 \mu m$ which roughly corresponds to half of the distance between the bottom and the top of the micro-channel indicating that flow visualisation is possible over the entire range of biomass concentrations explored.
The main tool we have used to quantitatively describe the flow fields was the Digital Particle Image Velocimetry (DPIV)  implemented in the house under \trademark{Matlab}, \cite{scarano, piv2}. For this purpose, a sequence of flow images has been acquired at half distance between the bottom and the top plate $z \approx H/2$ and a frequency of $30$ frames per second (fps) during the hydrostatic pressure ramp illustrated in Fig. \ref{fig:setup}(b).
Corresponding to low values of the driving pressure, several flow images have been skipped in order to increase the inter-frame and maintain the average displacement of the tracer particles in the optimal range of $5$ to $15$ pixels, \cite{piv2, scarano}. Together with this, the size of the smallest interrogation window has been adapted to the mean flow velocity corresponding to each step of the pressure ramp. 
The spatial resolution of the measured flow fields (as defined by the size of the smallest interrogation window and the interrogation window overlap) varied (depending on the signal to noise of the DPIV correlation)  between $3.75 \mu m$ and $7.5 \mu m$. 
The overall instrumental accuracy of our DPIV measurements (assessed via the standard deviation of the temporal fluctuations of the velocity) was about $4 \%$ of the mean flow speed in the bulk of the flow and about $6 \%$ near the boundary. Exceptionally, the accuracy was lower when very dilute micro-algae solutions were used at the largest driving pressure drops but it never exceeded $10 \%$ of the mean flow speed. This limitation of our measurements is solely related to the modest image acquisition speed. The flow fields have been averaged during the entire duration of each step of the controlled pressure ramp schematically illustrated in Fig. \ref{fig:setup}(b). This procedure equally allows one to estimate the level of fluctuations of the flow fields by calculating the corresponding root mean square deviation (rms). 

\subsection{Preparation of the \textit{Chlorella} micro-algae suspensions} \label{subsec:preparation}

We present below the main steps followed for the preparation of the micro-algae suspensions used through this study. For a more detailed description of the protocol the reader is referred to Ref. \cite{thesisAntoine}.
\textit{Chlorella} micro-algae were grown in a bubble column PBR.  The volume of the PBR was $0.12~m^3$ its diameter of $0.40~m$ and the specific illuminated area was $3.5~m^2/m^3$). The micro-algae cultures were illuminated with an incident photon flux density (PFD) of roughly $150~\mu mole_{h \nu} m^2~s^{-1}$ (white light). The culture medium was composed of (in g/l) $NH_4Cl, 1.45 $,  $MgSO_4.7H_2O, 0.28$, $CaCl_2.2H_2O, 0.05$, $KH_2PO_4, 0.61$, $NaHCO_3, 1.68$ and $0,5 ml~l^{-1}$ of Hutner solution (trace elements, see Ref. \cite{harris} for a complete description).
 Once collected, the micro-algal suspensions were artificially concentrated by means of centrifugation at $5000~g$ during $15$ minutes. The dry biomass concentration $C_x$ of the concentrate was then measured. Filters (glass micro-fiber filters, GF/F Whatman) with $0.22$ micrometer pore size were dried during $24$ hours in a $100^o C$ oven. They were then weighted and used to filter a known volume of algal suspension. The filters were finally placed for $24$ hours in a $100^o C$ oven and re-weighted. This duration was sufficient to obtain a complete drying. The weight difference was finally used to calculate the dry biomass concentration of the concentrate.
A range of samples of known concentration (here expressed in volume fraction, $\Phi_v$) was then prepared by diluting the concentrated wet suspension with the supernatant previously removed. The volume fraction ($\Phi_v$) was calculated from the dry weight concentration according to the equations below:

\begin{eqnarray}\label{eq:volumefraction}
\Phi_v=\frac{C_x}{\rho_W w_{D}} \\
%\rho_p=x_w \rho_{sec} + \left (  1-x_w \right) \rho_{sec}\\
\rho_W=x_W  \rho_{0} + \left ( 1-x_W \right ) \rho_D
\end{eqnarray}

Here $\rho_W$ and $\rho_{D}$ stand for the density of the wet and dry biomass, respectively and $\rho_0$ is density of the water and $w_D=\frac{\left( 1-x_W \right) \rho_D}{\rho_W}$ stands for the mass content in dry solid in the micro-algae and $x_W$ is the water content in wet micro-algae.
The water mass content $x_w$ and density of dry biomass $\rho_{D}$ necessary for the calculation were measured $0.82$ and $1400~kg m^{-3}$, respectively.

The relevant parameters of the suspensions used in this study are given in Table \ref{tab:suspensions}.
\begin{table}
\centering
\begin{tabular}{@{} l Lll L L  L@{} >{\kern\tabcolsep}l @{}}    \toprule
\rowcolor{white}[0pt][0pt] \textbf{Solution} & $\mathbf{C_x(g/l)}$ &$\mathbf{C_x^{rms}(g/l)}$& $\mathbf{\Phi_v}$&$\mathbf{\Phi_v^{rms}}$& \emph{\textbf{Behaviour}}   \\\midrule
1    & 4.54  & 0.17  & 0.02  & 0.0005 & \textbf{(N)}  \\ 
 2  & 29.74 & 1.15 & 0.12 & 0.004 & \textbf{(ST)}\\ 
 3  & 143.7 & 4.68 & 0.57 & 0.017 & \textbf{(YS)}\\\bottomrule
 \hline
 
\end{tabular}
\caption{Details of the \textit{Chlorella} suspensions used in the experiments. The superscripts \textit{"rms"} stand for the root mean square deviation.
The letters in the last column refer to the rheological behaviour of the suspension (see text for description): (N) - Newtonian, (ST) - shear thinning and (YS) - yield stress.}
\label{tab:suspensions}
\end{table}

Note the largest volume fraction we have investigated is relatively close to the maximum packing fraction corresponding to a maximally random jammed state, $\Phi_v = 0.637$ ($C_x = 159.25~g/l$).

\subsection{Rheological characterisation of the suspensions} \label{subsec:rheocharacterisation}

The rheological behaviour of the micro-algae solutions has been investigated using a Haake Mars $III$ rotational rheometer (Thermo Fisher Scientific, Germany). The rheometer is equipped with a nano-torque module which allows accurate measurements of dilute samples in a range of low applied stresses. 
The measurements have been performed with a cone-plate geometry (60 $mm$ radius and 2 deg truncation angle). Prior to each  test, the samples have been carefully loaded paying attention to the reproducibility of the shape of the free meniscus in subsequent experiments. To prevent the evaporation of the sample during long tests, the fluid meniscus has been covered by a protective thin film of silicone oil. Additionally, the entire geometry has been enclosed in a bell-shaped chamber made of a thermally insulating material (teflon). 
All the rheological tests have been performed at room temperature, $T= 20 ^oC$.
The Reynolds number was of order of unity (or smaller) during each of the rheological tests reported in the paper so inertial effects were absent.  The rheological behaviour of the micro-algae suspensions has been assessed by controlled stress flow ramps. To probe the reversibility of the deformation states, both increasing and decreasing stress ramps have been performed. Corresponding to each value of the applied stress, the rheological response of the suspension has been averaged during a finite time interval $t_0 = 5~s$.

\section{Results} \label{sec:results}

\subsection{Rheological properties of \textit{Chlorella} suspensions at various volume fractions} \label{subsec:rheo}

Prior to describing the microscopic flows of \textit{Chlorella} suspensions with various volume fractions $\Phi_v$ and for various driving pressure drops $
\Delta p$ we recall below several important rheological properties of the \textit{Chlorella} suspensions. For a more comprehensive and systematic discussion, the reader is referred to our recent rheological investigation, \cite{rheochlorella}.

\begin{figure*}
\centering
\subfigure[]{
     \includegraphics [width=7cm] {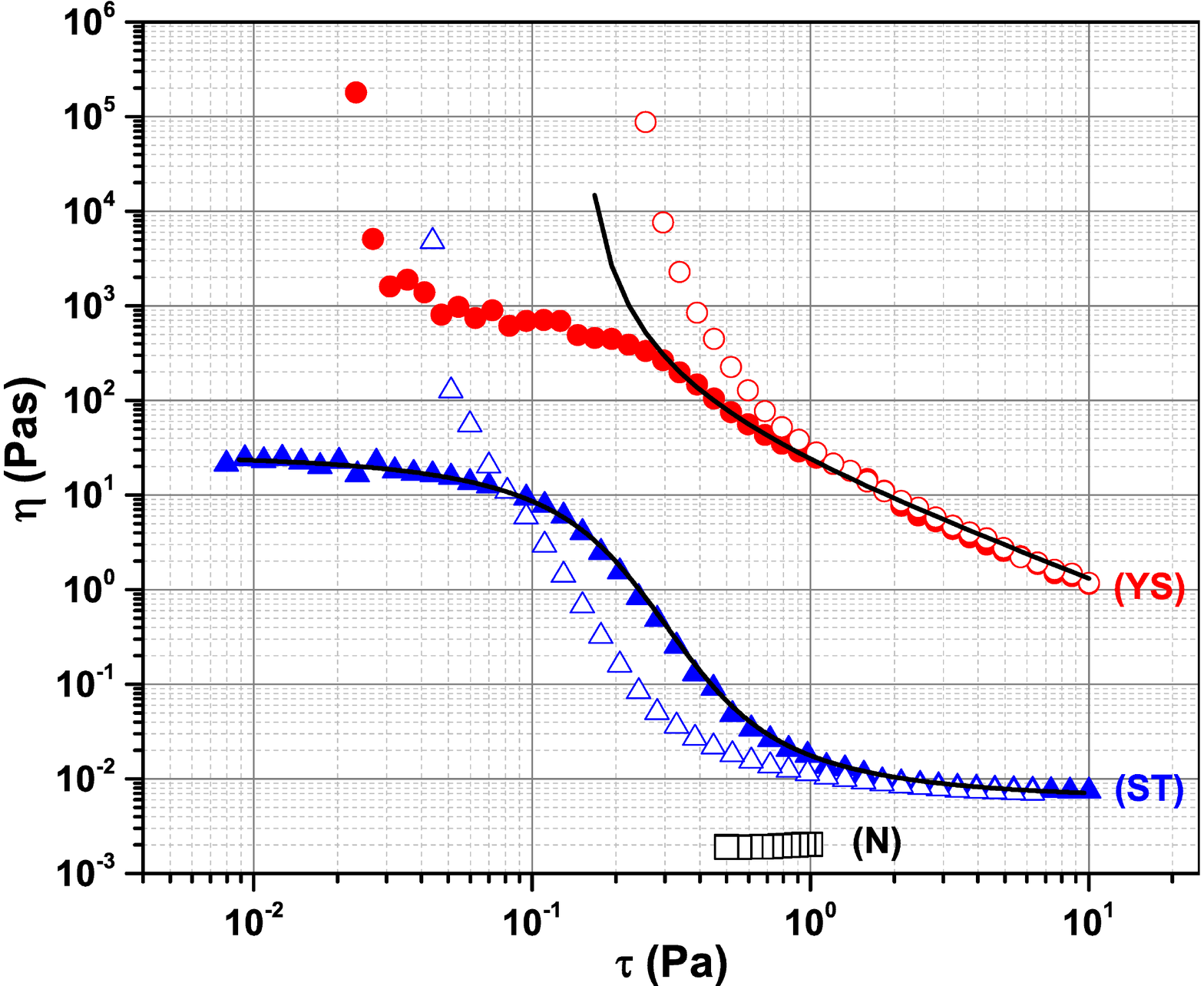}
     \label{sample_viscosities}
}
\subfigure[]{
                 \includegraphics [width=7cm] {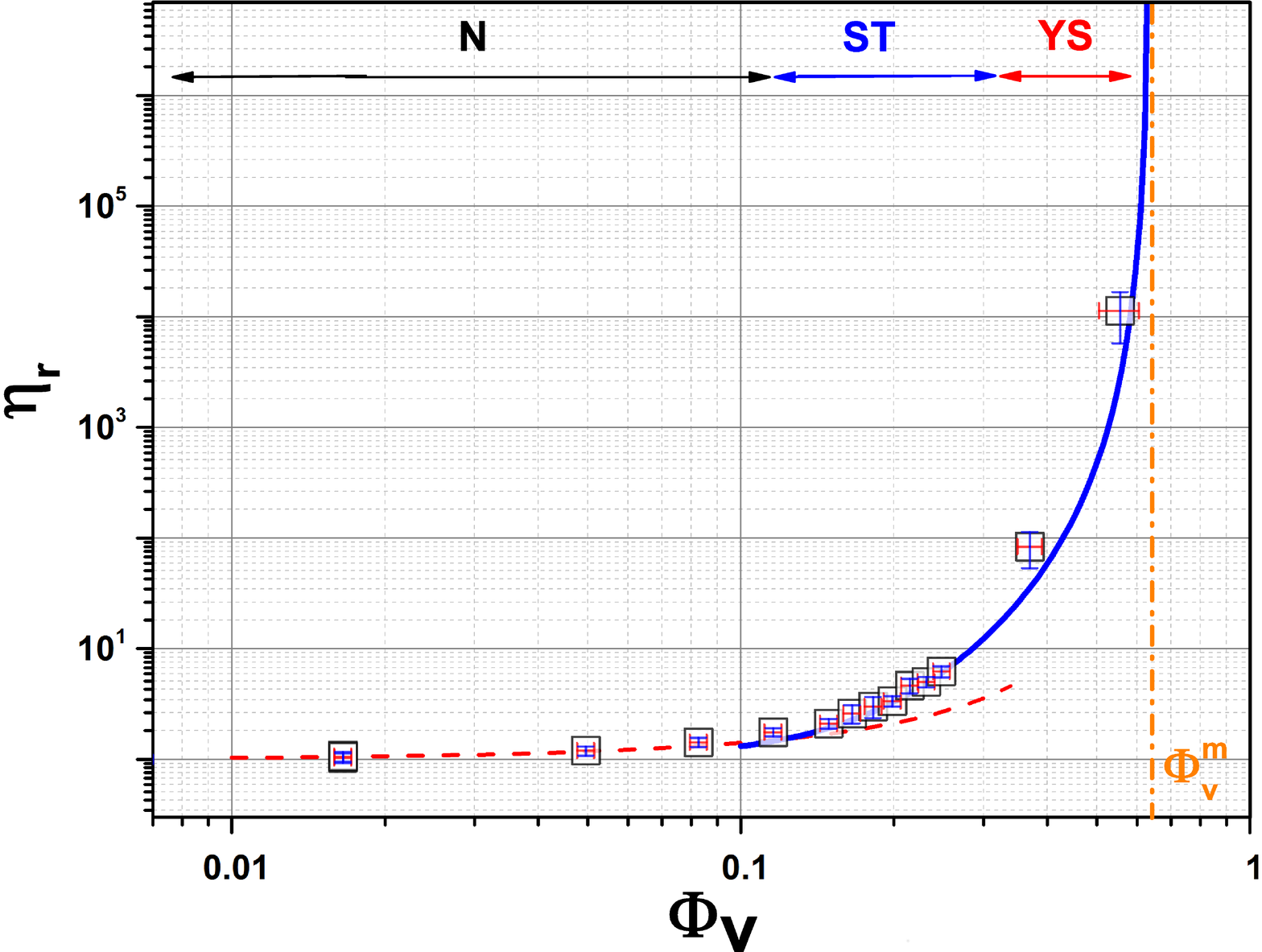}
    \label{relative_viscosity}}
    
\caption[Rheological characterisation of the \textit{Chlorella} suspensions]{\subref{sample_viscosities} Measurements of the apparent viscosity during an increasing/decreasing controlled stress ramp performed with suspensions at several volume fractions: $\textcolor{black}{(\blacksquare, \square)}$ - $\Phi_v= 0.08$, $\textcolor{blue}{(\blacktriangle, \bigtriangleup)}$ - $\Phi_v= 0.12$,   $\textcolor{red}{(\bullet, \circ)}$ - $\Phi_v= 0.57$. The full/empty symbols refer to the increasing/decreasing branches of the stress ramp, respectively. The full lines are nonlinear fitting functions (see the text for the explanation). The labels indicate the experimentally observed
rheological regimes: \textbf{N} - Newtonian, \textbf{ST} - shear
thinning, \textbf{YS} - yield stress.  \subref{relative_viscosity}  Dependence of the relative viscosity of the micro-algae suspension $\eta_r$ on the volume fraction of the microalgue $\Phi_V
$. The dashed line is a fit according to the Quemada model
\cite{quemada1} and the full line is a fit according to the Simha
model, \cite{simha}. The labels indicate the experimentally observed
rheological regimes: \textbf{N} - Newtonian, \textbf{ST} - shear
thinning, \textbf{YS} - yield stress. Within the \textbf{ST} and
\textbf{YS} regimes, the relative viscosity has been measured at a
fixed stress, $\tau = 1Pa$. }
\label{fig:rheology}
\end{figure*}

 The dependence of the apparent shear viscosity $\eta$ on the applied stress $\tau$ measured for three distinct volume fractions $\Phi_v$ (see Table \ref{tab:suspensions} for details) is presented in Fig. \ref{sample_viscosities}.  The full/empty symbols refer to the data acquired on the increasing/decreasing branch of the controlled stress ramp respectively. For the lowest volume fraction investigated, $\Phi_v= 0.08$, the viscosity is independent on the applied stress and a Newtonian behaviour is observed (the squares in Fig. \ref{sample_viscosities}). As the data acquired for increasing/decreasing stresses overlap nearly perfectly, one can conclude that no thixotropic effects are present.

A fundamentally different rheological behaviour is observed at larger volume fractions, $\Phi_v = 0.12$. Corresponding to the increasing branch of the stress ramp (the full triangles in Fig. \ref{sample_viscosities}) we observe a shear thinning rheological behaviour that can be well described by the Cross model. Based on in-situ observation of individual \textit{Chlorella} cells during the rheological tests, we have shown in Ref. \cite{rheochlorella} that this behaviour is related to the presence of flocs of cells. As the applied stresses are increased,  these flocs are gradually destroyed which translates into a decrease of the apparent viscosity from a first plateau visible at low applied stresses down to a (second) plateau at high applied stresses. Unlike for the Newtonian regime observed at low volume fractions, the intermediate deformation states are no longer recovered upon increasing/decreasing the applied stresses and a rheological hysteresis is observed. As the applied stresses are decreased, the individual cells re-aggregate at a faster rate than the initial cell flocs were destroyed along the increasing branch of the stress ramp which ultimately translates into an apparent viscosity divergence at low applied stresses. 
Corresponding to the largest volume fraction investigated through our microfluidic experiments $\Phi_v=0.57$, a divergence of the apparent viscosity in a range of low applied stresses is observed on the increasing branch of the controlled stress ramp (the full circles in Fig. \ref{sample_viscosities}). This is consistent with a yield stress behaviour. By direct visualisation of individual \textit{Chlorella} cells within the gap of the cone and plate geometry, we have shown in Ref.  \cite{rheochlorella} that the yield stress behaviour is related to the presence of large scale aggregates of cells (reaching $100 \mu m$ in size), see Fig. 7 in Ref.  \cite{rheochlorella}. At large applied stresses the apparent viscosity measurements may by formally fitted by the Herschel - Bulkley model, \cite{hb}, which gives an estimate for the yield stress $\tau_y = 0.47 \pm 0.1$. In a range of low applied stresses (around the yield point), however, the rheological behaviour significantly departs from the Herschel - Bulkley model. We note that the deformation states are reversible only after the suspension is fully yielded, i.e. at large applied stresses. 

The various rheological regimes observed when the volume fraction $\Phi_v$ is increased briefly described above may be summarised by plotting the
dependence of the relative effective viscosity $\eta_r = \frac{\eta}
{\eta_0}$ of \textit{Chlorella} micro-algae suspensions measured at a
constant applied stress $\tau=1~Pa$ on the volume fraction $\Phi_v$,
Fig. \ref{relative_viscosity}. Here $\eta_0 \approx 2~mPas$ stands for
the viscosity of the culture medium. The value of the applied stress is chosen such as the rheological measurements are reversible upon increasing/decreasing stresses. We have verified  that choosing any stress value within this range would lead to the same qualitative behaviour. 

Within the Newtonian volume fraction regime observed for dilute
\textit{Chlorella} suspensions (N) the volume fraction dependence of the relative viscosity $\eta_r$ is often modelled by the Quemada model, \cite{quemada1}:

\begin{equation} \label{eq:quemada}
\eta_r =\left (1 - \frac{\Phi_v}{\Phi_v^m} \right ) ^{-2}
\end{equation}

The equation above with $\Phi_v^m=0.637$ can reliably describe the viscosity measurements in the range $\Phi_v \leq 0.115$ (the dotted
line in Fig. \ref{relative_viscosity}).

Within the concentrated and the highly concentrated regimes (ST, YS) the volume fraction dependence of the
relative viscosity is no longer accurately predicted by the Quemada model.
According to the Simha's cellular model  \cite{simha} which has proven its ability to describe the rheological behaviour of the
highly concentrated suspensions, the relative viscosity of the suspension should satisfy:

\begin{equation} \label{eq:simha}
\lim _{\Phi_v \to \Phi_v^m}{\eta_r} =1+ \frac{54}{5} f^3
\frac{\Phi_v^2}{\left (1-\frac{\Phi_v}{\Phi_v^m} \right )^3}
\end{equation}
where $f$ is a semi-empirical parameter of order of unity.

The dependence of the relative viscosity measured at a constant applied stress $\tau=1Pa$ within the ST and YS
regimes can be fitted by Eq. \ref{eq:simha} with $f=1.2$ and $\Phi_v^m=0.637$ (the full line in Fig. \ref{relative_viscosity}).
This value of the maximum packing fraction $\Phi_v^m$ matches closely the theoretical prediction for a \textit{maximally random
jammed state}.

\subsection{Microscopic flow behaviour of \textit{Chlorella} suspensions at various volume fractions} \label{subsec:micro}

We focus in the following on the experimental characterisation of the microscopic flow profiles of \textit{Chlorella} suspensions with various concentrations driven at several pressure drops. The measurements have been performed at a position located roughly $2~cm$ downstream. As the Reynolds numbers are very small (see the discussion presented in Sec. \ref{subsec:microflow}), the laminar flow is fully developed at this position.   

\begin{figure}
\centering
  \includegraphics[width=0.4\textwidth]{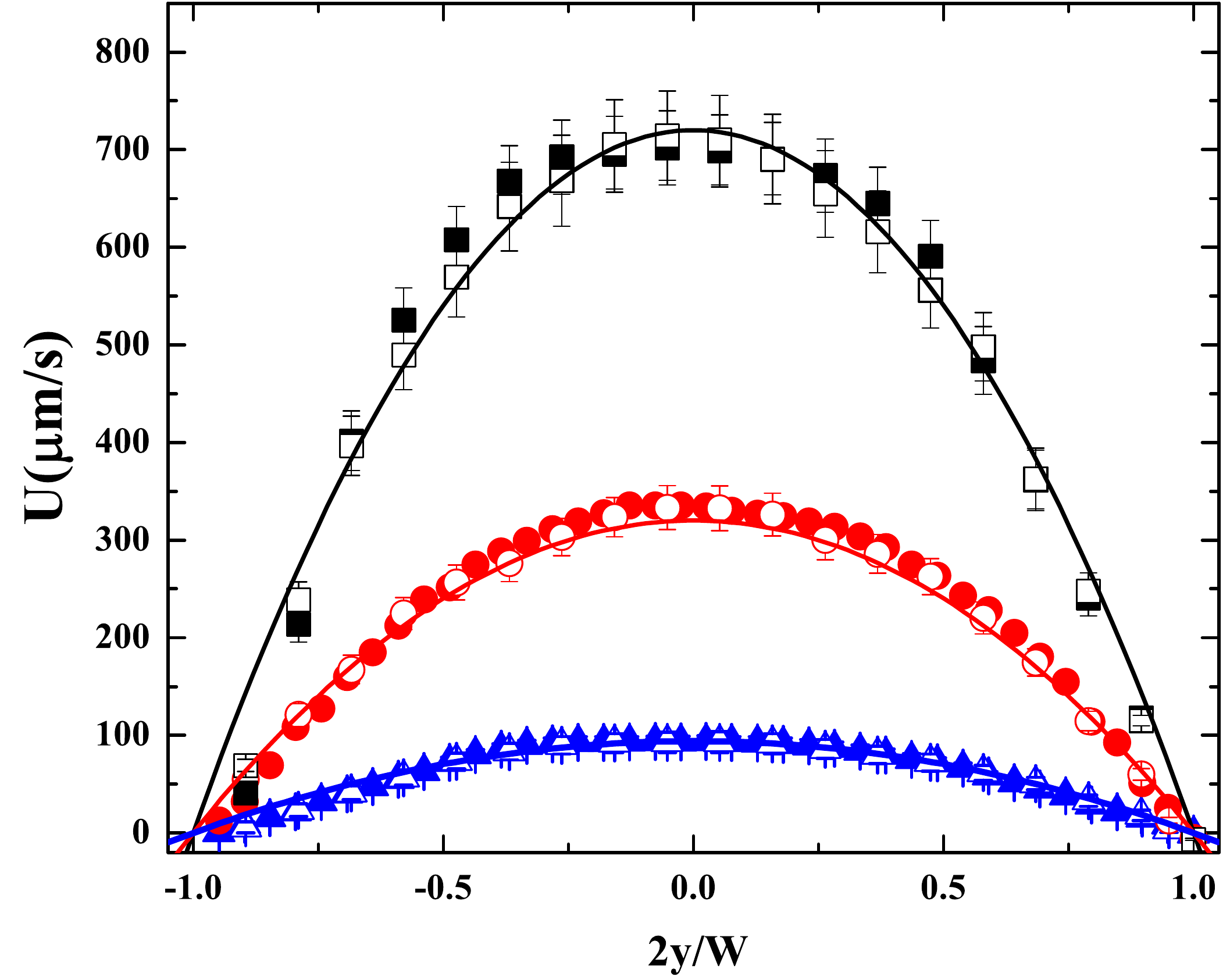}
\caption{   Time averaged velocity profiles measured with a $\Phi_v = 0.02$ suspension at various pressure drops:
$\Delta p=8.3 Pa$ ($\textcolor{blue}{\blacktriangle,
\bigtriangleup}$), $\Delta p=25 Pa$ ($\textcolor{red}{\bullet, \circ}$), $\Delta p=50 Pa$ ($\blacksquare$, $\square$). The full/empty symbols refer to the increasing/decreasing branch of the pressure ramp, respectively.}
\label{fig:profilN}
\end{figure}

Measurements of the time averaged (over $t_0=6~s$) transversal 
profiles of the axial velocity $U = \left < u(y, t) \right >_{ t}$  performed with a very dilute \textit{Chlorella} suspension ($\Phi_v=0.02$) for several values of the driving pressure drop $\Delta p$ on both the increasing and the decreasing branch of the pressure ramp are presented in Fig.
\ref{fig:profilN}. 

The analytical solution of a low Reynolds number laminar flow of a
Newtonian fluid in a rectangular channel is:

\begin{equation}\label{eq:channelflow}
U(y,z=H/2) =-\frac{4 H^2 \Delta p}{\pi^3 \eta L} \sum_{n, odd}
^{\infty} \frac{1}{n^3} \left [ 1 - \frac{cosh \left(\frac{n\pi y}
{H} \right)}{cosh \left(\frac{n\pi W}{H} \right)}\right]
\end{equation}

Within this dilute regime and regardless the value of the driving pressure drop $\Delta p$ the velocity profiles are consistent with a Poiseuille flow and may be fitted by Eq. \ref{eq:channelflow}, the full lines in Fig. \ref{fig:profilN}. As the series involved in the right hand side of Eq. \ref{eq:channelflow} converges rather fast, retaining only three terms was sufficient to obtain a reliable fit. 
The velocity profiles measured on the increasing/decreasing branches of the controlled pressure ramp coincide indicating that the flow states are fully reversible. This finding is consistent with the reversibility of the rheological flow curves upon increasing/decreasing stresses, the squares in Fig. \ref{sample_viscosities}.

We note that the apparent velocity slip at the wall is negligible and solely related to the instrumental error of the micro-PIV measurements (the full lines in
Fig. \ref{fig:profilN}). This observation is consistent with the visual assessment of the wall slip performed in a rheometric flow for low values of the cell volume fraction within the Newtonian regime N, \cite{rheochlorella}.

\begin{figure}
\centering
  \includegraphics[width=0.4\textwidth]{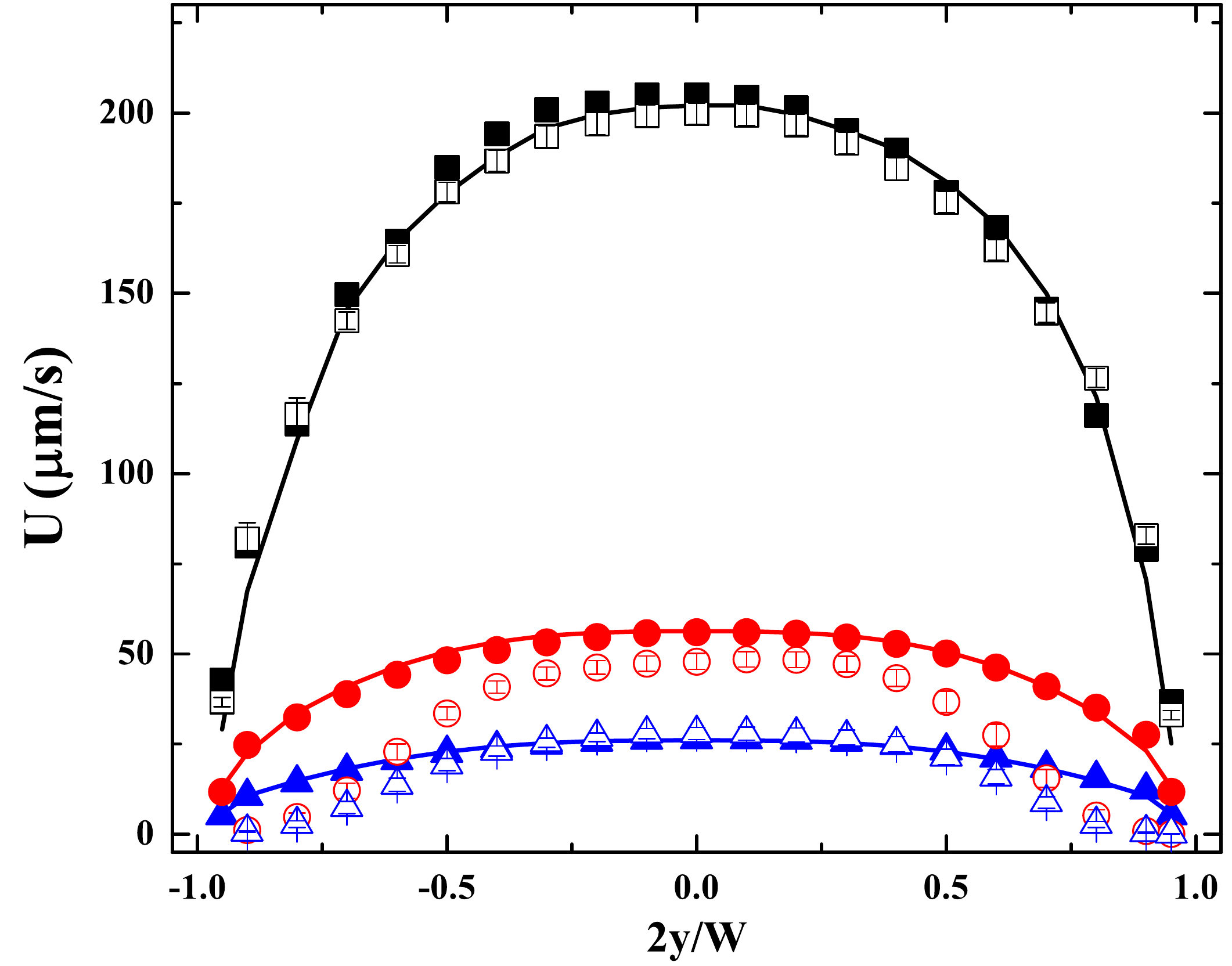}
\caption{ Time averaged velocity profiles measured with a $\Phi_v = 0.12$ suspension at various pressure drops:
$\Delta p=10 Pa$ ($\textcolor{blue}{\blacktriangle, \bigtriangleup}$), $\Delta p=16.7
Pa$ ($\textcolor{red}{\bullet, \circ}$), $\Delta p=46.7 Pa$ ($\blacksquare$, $\square$ ). The full/empty symbols refer to the increasing/decreasing branch of the pressure ramp, respectively.}
\label{fig:profilST}
\end{figure}

Similar measurements of the time averaged transversal profiles of the axial velcity performed for several driving pressures with mildly concentrated suspension ($\Phi_v = 0.12$) corresponding to the shear thinning rheological regime (see the triangles in Fig. \ref{sample_viscosities}) are presented in Fig. \ref{fig:profilST}. 

Corresponding to the largest values of the driving pressure,  $\Delta p =46.7~Pa$ (the squares in Fig. \ref{fig:profilST}) the velocity profiles are no longer accurately described by a classical Poiseuille correlation as given by Eq. \ref{eq:channelflow} but by an Ostwald de Waele type equation with slip, \cite{schechter}:

\begin{equation}\label{eq:ostwald}
U=U_s+U_0 \left[ 1- \left \vert \frac{2y}{W} \right \vert  \right ]^{\frac{n+1}{n}}
\end{equation}

Here $U_s$ stands for the slip velocity which, unlike in the Newtonian case, is positive and significantly larger than the magnitude of the error bars. 
At this driving pressure the flow profiles are reversible upon increasing/decreasing pressure drops, see the full/empty squares in Fig. \ref{fig:profilST}. This corresponds to the reversible rheological regime visible in Fig. \ref{sample_viscosities} (the triangles) in the range $\tau \geq 3~Pa$.

\begin{figure}
\centering
  \includegraphics[width=0.4\textwidth]{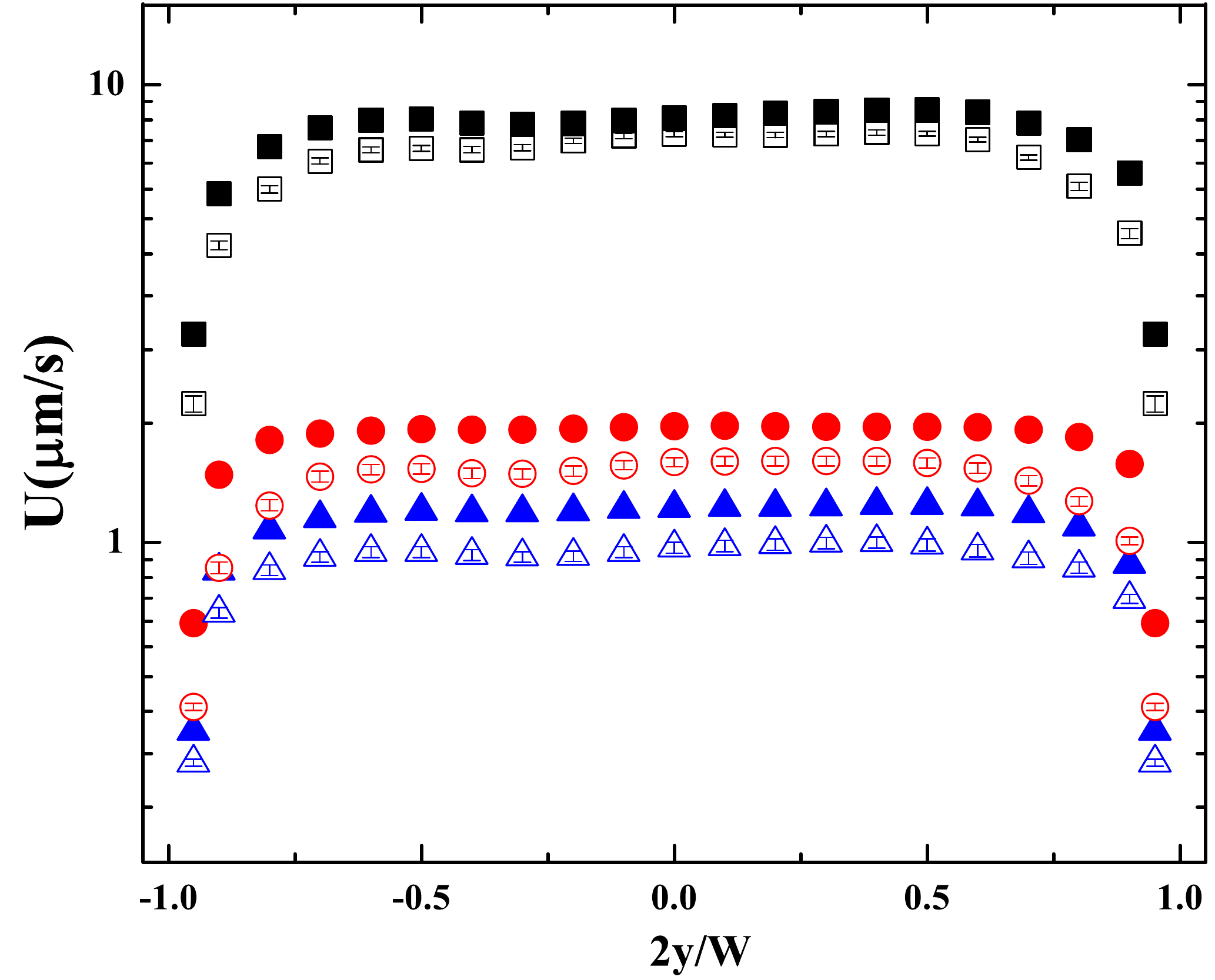}
\caption{  Time averaged velocity profiles measured with a $\Phi_v = 0.57$ suspension at various pressure drops:
$\Delta p=25 Pa$ ($\textcolor{blue}{\blacktriangle, \bigtriangleup}
$), $\Delta p=41.7 Pa$ ($\textcolor{red}{\bullet, \circ}$), $\Delta p=116.8 Pa$ ($\blacksquare$, $\square$). The full/empty symbols refer to the increasing/decreasing branch of the pressure ramp, respectively.}
\label{fig:profilYS}
\end{figure}

Corresponding to lower values of the driving pressure (the circles and the triangles in Fig. \ref{fig:profilST}) the hydrodynamic reversibility is broken: the profiles measured on the increasing/decreasing branches of the pressure ramp no longer coincide, particularly in the vicinity of the channel boundaries. This corroborates with the observation of a rheological hysteresis during the controlled stress ramp in the range $\tau < 3~Pa$. We also note that the velocity profiles acquired on the increasing branch of the controlled pressure ramps (the full circles and triangles in Fig. \ref{fig:profilST})) may be still fitted by Eq. \ref{eq:ostwald}.

Measurements of the time averaged transversal profiles of the axial velocity performed with the highly concentrated suspension with $\Phi_v = 0.57$ at three values of  the driving pressure drop are presented in Fig. \ref{fig:profilYS}. 
Regardless the value of the pressure drop, the velocity profiles exhibit a central plug region typically observed during viscoplastic flows and wall slip.  
As in the shear thinning case, the flow states are reversible only for the largest driving pressure drop, the squares in Fig. \ref{fig:profilYS}. The flow profiles acquired at lower driving pressures are, as in the shear thinning case, irreversible upon increasing/decreasing pressures, (the circles and the triangles in Fig. \ref{fig:profilYS}). This hydrodynamic irreversibility is equally related to the rheological hysteresis observed in Fig. \ref{sample_viscosities} (the circles) in the range $\tau \geq 1 Pa$.

\begin{figure}
\centering
  \includegraphics[width=0.6\textwidth]{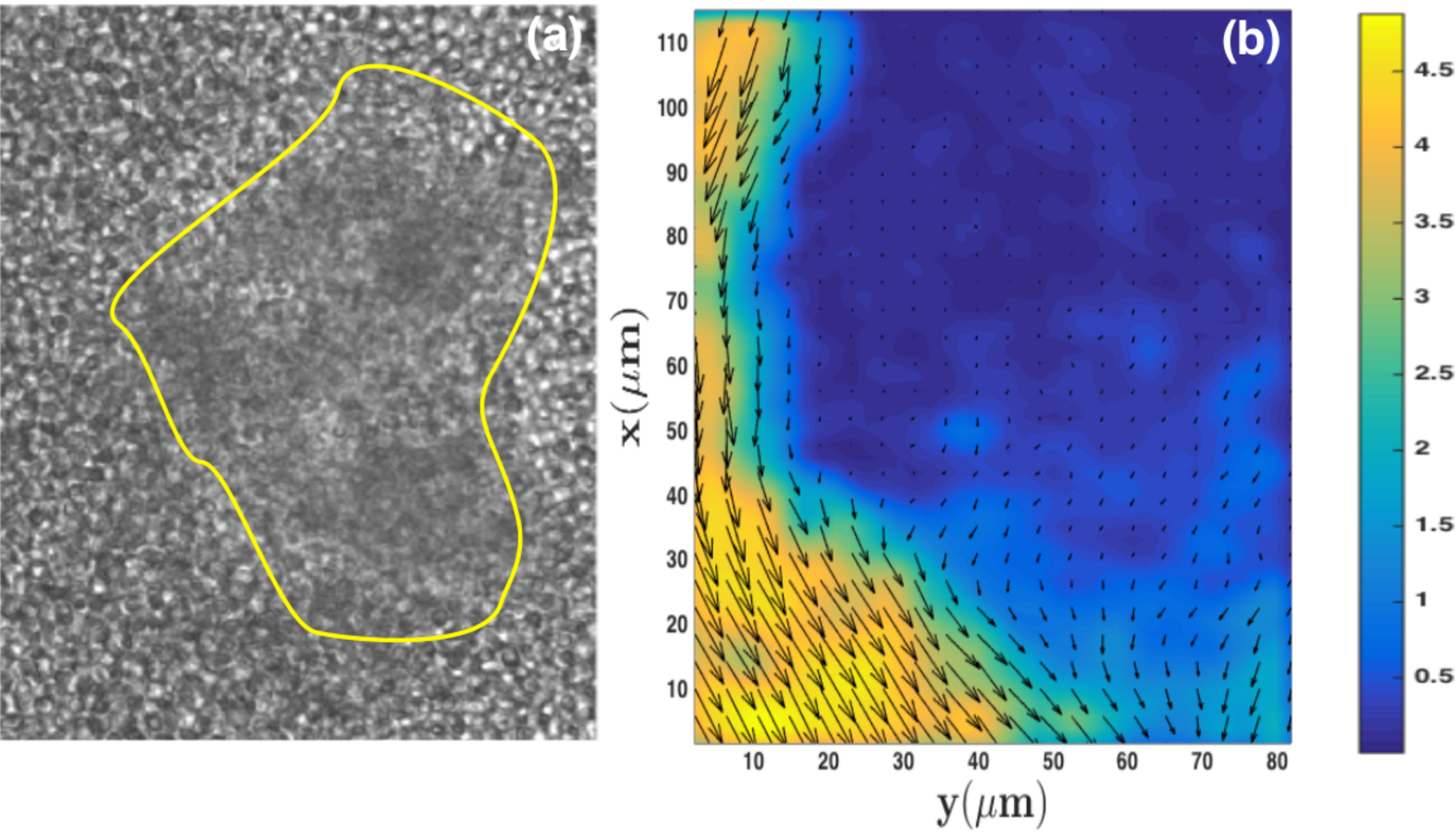}
\caption{ \textbf{(a)} Bright field micrograph of a concentrated \textit{Chlorella} suspension with $\Phi_v=0.57$ flowing in a microchannel. The driving presssure was $\Delta p = 41.7 Pa$. The closed curve highlights a large agglomerate of micro-algae. \textbf{(b)}  Flow field averaged during $t_0=6~s$. The right edge of the field of view corresponds to a channel boundary and the left one to the centreline of the channel. The colormap relates to the magnitude of the flow velocity. For clarity of the presentation, the vector field has been down-sampled by a factor of $2$ along each axis.}
\label{fig:plugfield}
\end{figure}

To gain further insights into the physical origins of the yield stress behaviour observed through the experiments performed with the concentrated suspensions, we present in Fig. \ref{fig:plugfield}(a) a bright field micrograph acquired during the flow of a  $\Phi_v = 0.57$ suspension at a driving pressure $\Delta p = 41.7~Pa$. The micrograph was acquired at a downstream position that was several $mm$ closer to the inlet than the position where the velocity profiles presented in Figs. \ref{fig:profilN}, \ref{fig:profilST}, \ref{fig:profilYS} were measured. One can observe a static large aggregate (nearly $60~\mu m$ in size) of \textit{Chlorella} cells formed in the vicinity of the flow channels's wall (the highlighted region in Fig. \ref{fig:plugfield}(a)). Whereas around this large aggregate one can still clearly distinguish individual \textit{Chlorella} cells,  much fewer details can be distinguished within the aggregate. This indicates that the cells forming the aggregate are the smallest ones which, at this volume fraction and optical magnification, makes them indistinguishable.  

The corresponding velocity field averaged during $t_0=6~s$ confirms the presence of the solid plug in the flow, Fig. \ref{fig:plugfield}(b)). One can conclude at this point that, as in the case of the rheometric flow (see Fig. 7 in Ref. \cite{rheochlorella}), the yield stress behaviour originates from the aggregation of individual cells in solid structures with sizes comparable to the size of the flow channel.

A first insight into the physics behind the formation of such large scale aggregates of small \textit{Chlorella} cells was given in Ref. \cite{rheochlorella} in terms of the superficial electrical charge of the individual cells which has been observed experimentally, \cite{kumar}. By identifying the centroids and contours of each individual cells and subsequent Voronoi triangulation analysis which allowed one to  correlate the spatial distribution of cells with their sizes we have concluded that the largest cells (carrying the largest superficial charge) repel the smaller ones forcing them to organise themselves in large scale aggregates, see Fig. 5 in Ref. \cite{rheochlorella}. Another factor that may promote the aggregation of individual cells may be related to the secretion of polysacharides \cite{Hadji}.

\begin{figure}
\centering
  \includegraphics[width=6 cm]{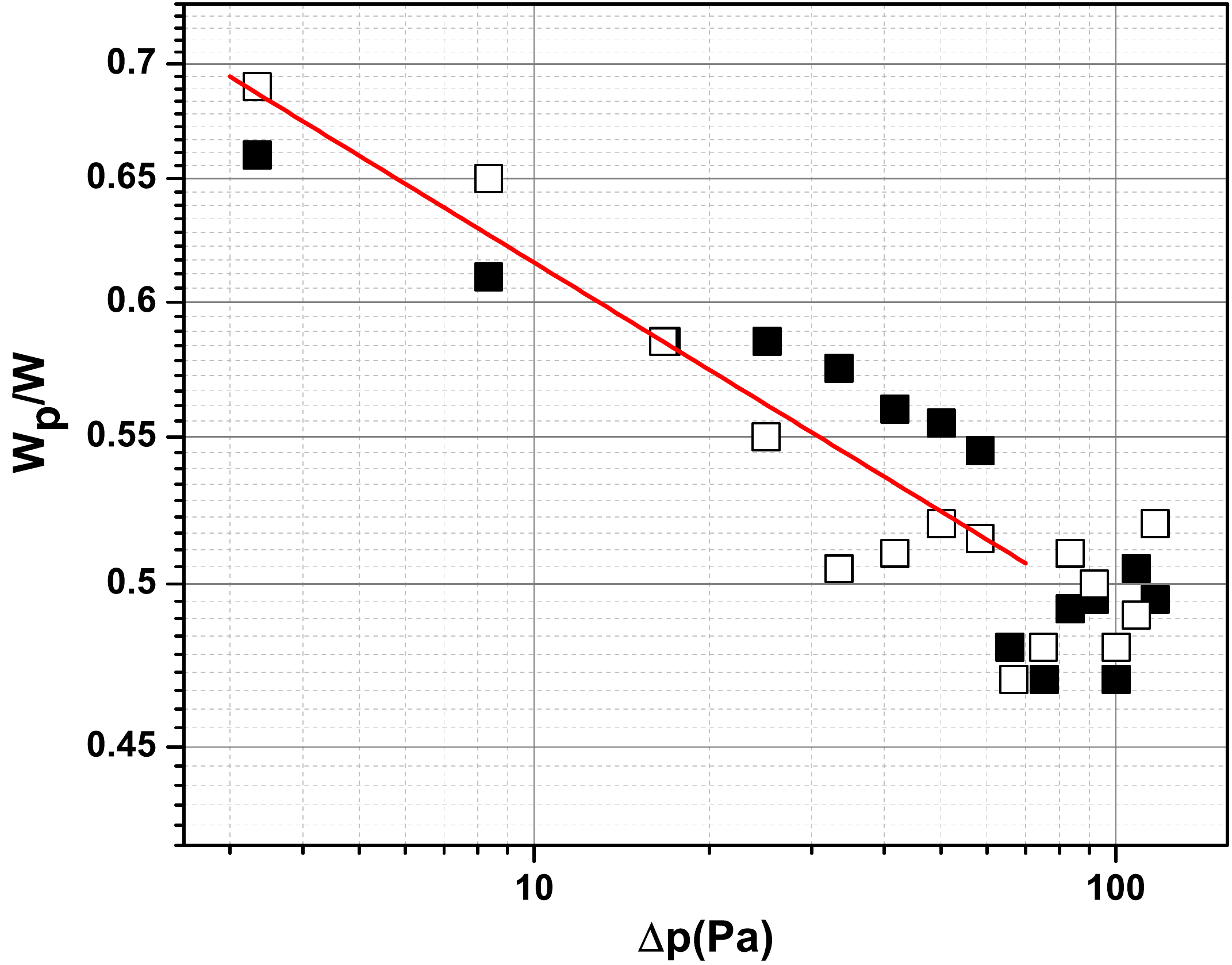}
\caption{ Dependence of the normalised size of the plug $W_p/W$ on the driving pressure drop measured for a suspension with $\Phi_v = 0.57$. The full/empty symbols distinguish between the data acquired on the increasing/decreasing branch of the stepped pressure
ramp, respectively. The full line is a guide for the eye, $W_p/W \propto \Delta p ^{-0.1}$.}
\label{fig:plugsize}
\end{figure}

The evolution of the size of the rigid plug $W_p$ defined by the extent of the flat central region of the time averaged velocity profiles (see Fig. \ref{fig:profilYS}) on the driving pressure droop is illustrated in Fig. \ref{fig:plugsize}. Whereas at low driving pressures the plug size decreases gradually with increasing pressures as one would expect for a gradual yielding process, a plateau is observed in the range $\Delta p \geq 70~Pa$. By analysis of individual flow images similar to that illustrated in Fig. \ref{fig:plugfield}(a) we could confirm that an aggregate of cells with an average size accounting for roughly $46 \%$ of the width of the channel "survives" even at highest value of the pressure drop we have explored. It is also interesting to note that before reaching the plateau the normalised width of the plug scales as $W_p/W \propto \Delta p ^{-0.1}$. This scaling is roughly an order of magnitude slower than that observed for a Carbopol gel, \cite{coussotbook,pipesteady,Poumaere201428}, $W_p/W \propto \Delta p ^{-1}$. 

\begin{figure*}[ht]
\centering
\subfigure[]{
     \includegraphics [width=6cm] {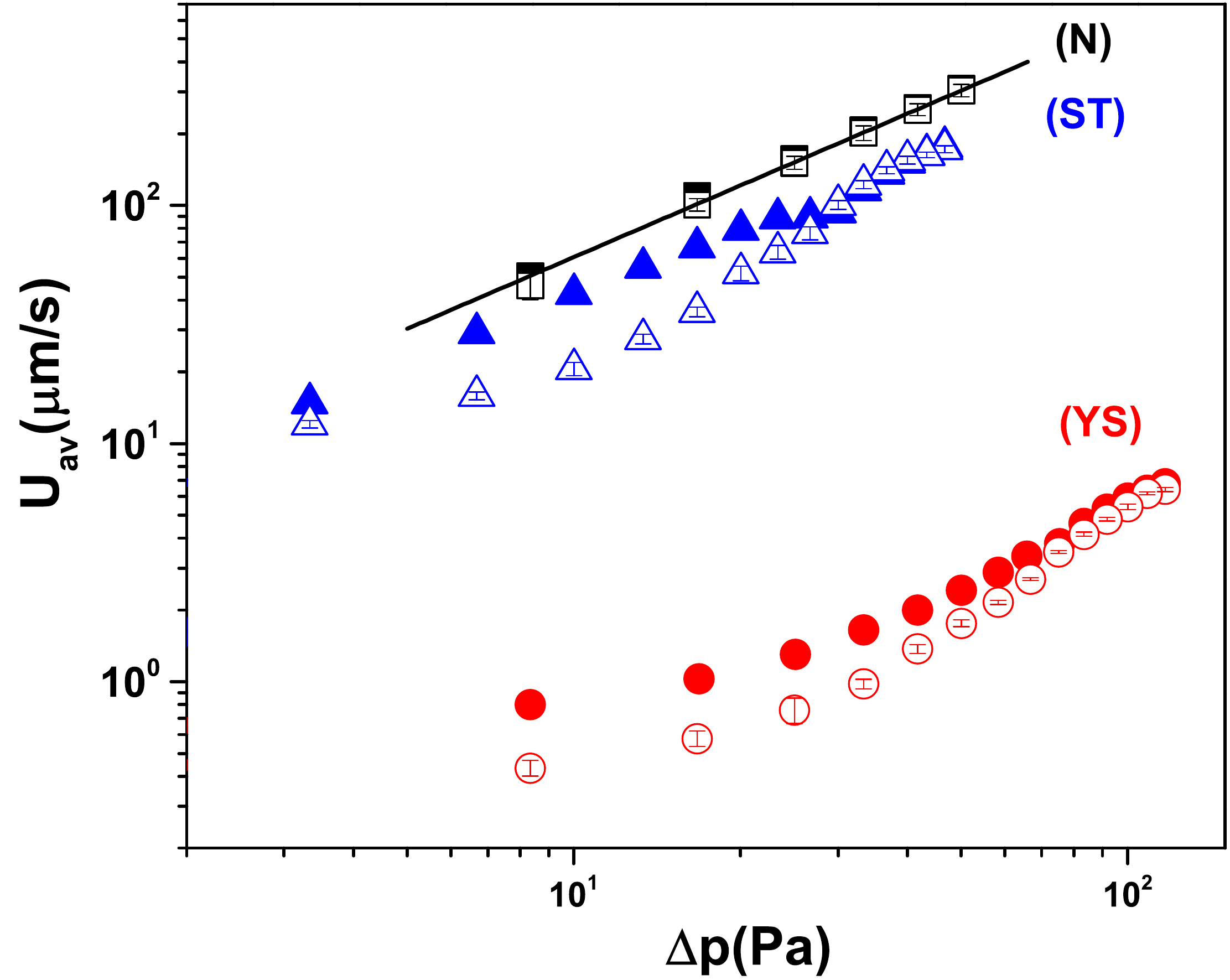}
     \label{fig:averagevelocity}
}
\subfigure[]{
                 \includegraphics [width=6cm] {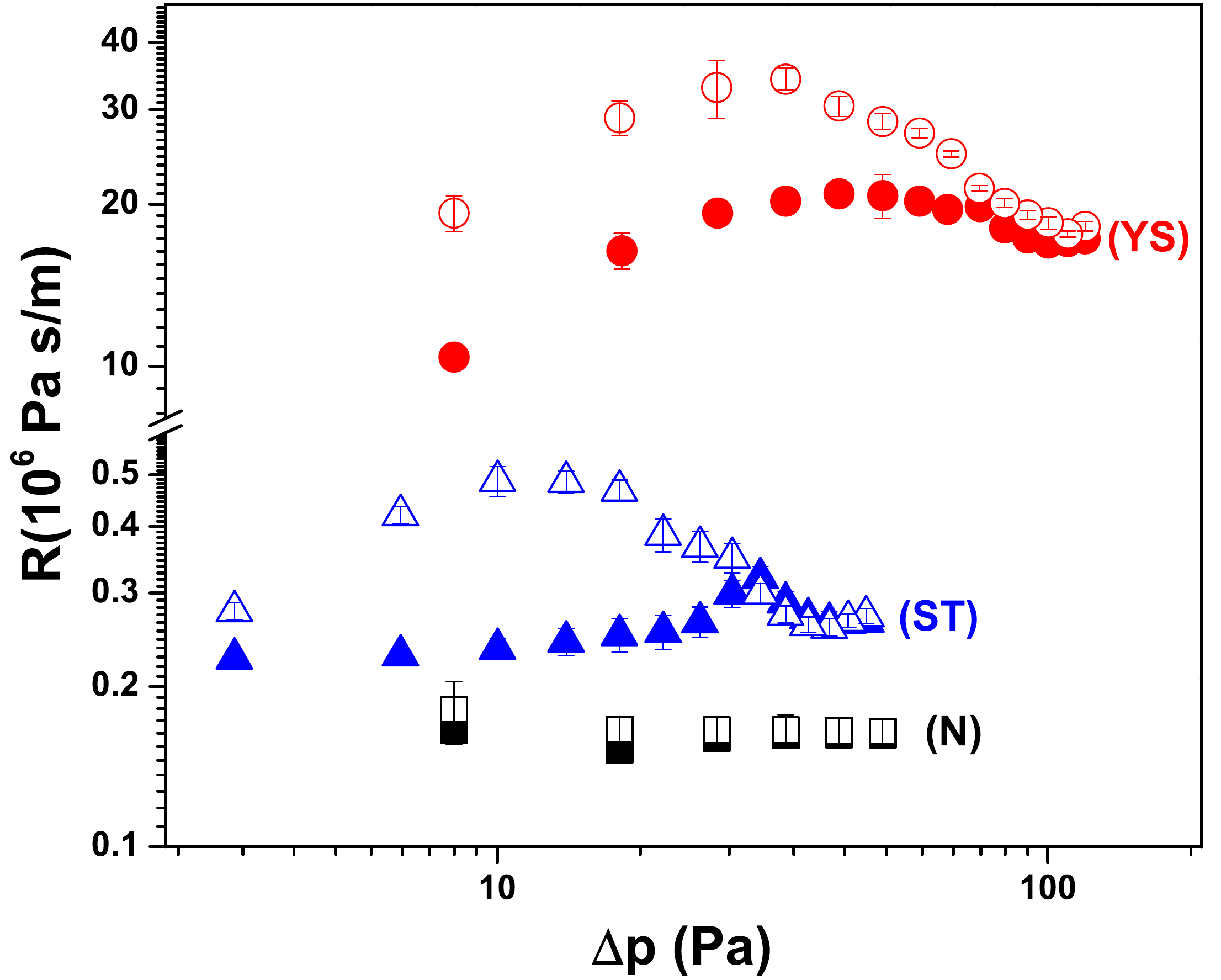}
    \label{fig:flowresistance}}
    
\caption[Dependence of the mean flow speed and of the flow resistance on the pressure drop.]{\subref{fig:averagevelocity} The dependence of the average flow speed $U_{av}$ on the
driving pressure drop $\Delta p$ measured for \textit{Chlorella}
suspensions at various volume fractions: ($\blacksquare$, $\square$
) - $\Phi_v=0.02$, ($\textcolor{blue}{\blacktriangle,
\bigtriangleup}$) - $\Phi_v=0.12$,  ($\textcolor{red}
{\bullet, \circ}$) - $\Phi_v=0.57$. The labels \textbf{(N)}, \textbf{(ST)},
\textbf{(YS)} stand for Newtonian, shear thinning and yield stress,
respectively. The full/empty symbols distinguish between the data
acquired on the increasing/decreasing branch of the stepped pressure
ramp, respectively. The full line is a linear fit.  \subref{fig:flowresistance}  Dependence of the flow resistance $R$ on the driving pressure drop $\Delta p$. The symbols are the same as in panel  \subref{fig:averagevelocity}.}
\label{fig:meanflow}
\end{figure*}

The dependencies of the flow velocity averaged in both time and along the $x$ direction $U_{av} = \left < u(\vec{r}, t) \right >_{\vec{r}, t}$ on the driving pressure drop $\Delta p$ are presented in Fig.  \ref{fig:averagevelocity}. For the lowest volume fraction corresponding to the Newtonian rheological regime  (the full/empty squares) the dependence is linear which is consistent with the Poieseuille type flow profiles illustrated in Fig.  \ref{fig:profilN}. The slope of the linear dependence is consistent with the Newtonian viscosity of the suspension. The measurements performed on the increasing/decreasing branches of the controlled pressure ramp (the full/empty symbols) overlap which demonstrates the hydrodynamic reversibility of the flow. This is consistent with the lack of a rheological hysteresis illustrated in Fig. \ref{sample_viscosities} (the squares) within this volume fraction regime. 

For the intermediate value of the volume fraction $\Phi_v = 0.12$ corresponding to the shear thinning rheological regime the dependence of the mean flow speed on the driving pressure is no longer linear (the triangles in Fig.  \ref{fig:averagevelocity}). Upon increasing/decreasing driving pressures the hydrodynamic reversibility is preserved only at large pressure drops, $\Delta p \geq 30~Pa$. The irreversible flow behaviour observed for low values of the driving pressure is also fully consistent with the rheological hysteresis illustrated in Fig. \ref{sample_viscosities}.

As the volume fraction is increased even further within the yield stress regime $\Phi_v=0.57$ a similar hydrodynamic irreversibility is observed in a range of small driving pressures. Like in the shear thinning case, the hydrodynamic reversibility is restored only at larger driving pressures, $\Delta p >30~Pa$. 
 
The emergence of the hysteresis of the flow states observed within the shear thinning and the yield stress regimes is typically associated to a thixotropic bulk rheological behaviour and observed in various other complex fluids (micellar solutions, physical and colloidal gels) subjected to an external forcing. 
 Thus, these experimental observations are of a broader fundamental interest and deserve a separate discussion. As already discussed above, the \textit{Chlorella} cells are superficially charged which leads to formation of aggregates via electrostatic repulsions. In the presence of a time varying external forcing such as the hydrostatic pressure employed in our experiments (Fig. \ref{fig:setup}(b)), the dynamics of the system is governed by a competition between structuring of the microscopic constituents and their break-up triggered by the local shear. 
 It has been recently shown that a plausible cause underlying the emergence irreversibility is related to the presence of interactions among microscopic constituents, \cite{gibbs,gibbs1}. Thus it is shown by both microscopic Gibbs field simulations in \cite{gibbs} and by a nonlinear dynamical system approach in \cite{gibbs1}  that in the presence of weak interactions among the microscopic constituents (e.g. the case of a Carbopol gel, \cite{solidfluid}) the deformation states are recoverable upon increasing/decreasing forcing only in the asymptotic limit of a steady state forcing (the applied stress is maintained constant for a period of time $t_0$ significantly longer than any characteristic time associated to the microstructure). On the other hand, if the magnitude of the interactions exceeds a critical threshold, a hysteresis of  the deformation states will \textit{always} be observed (e. g. the case of laponite, bentonite, carbon black suspensions). In a qualitative agreement with these theoretical predictions, in the case of \textit{Chlorella} suspensions with concentrations within the range of the shear thinning a yield stress regimes, due to the interactions among neighbouring cells, a rheological hysteresis is observed in range of small applied stresses even during steady state flow ramps.

From an energetic standpoint it is instructive to monitor the dependence of the flow resistance defined as $R=\frac{\Delta p}{U_{av}}$ on the driving pressure drop, Fig. \ref{fig:flowresistance}. An increase of the cell volume fraction from the Newtonian regime to the yield stress regime is accompanied by a sharp increase of the flow resistance which, at moderate driving pressure drops, may reach a value $220$ times larger than the Newtonian value. This indicates that efficient stirring of concentrated suspensions of micro-algae in intensified PBR's comes at a non negligible energetic cost that should be accounted for in the context of optimisation of the micro-algae production. 

We present in the following a more in depth analysis of the wall slip behaviour observed within the shear thinning and yield stress volume fraction regimes illustrated in Figs. \ref{fig:profilST}, \ref{fig:profilYS}. The slip behaviour may be described by focusing on the dependence of the slip velocity $U_s$ on the wall shear stress $\tau_w$. 

\begin{figure*}[ht]
\centering
\subfigure[]{
     \includegraphics [width=5.45cm] {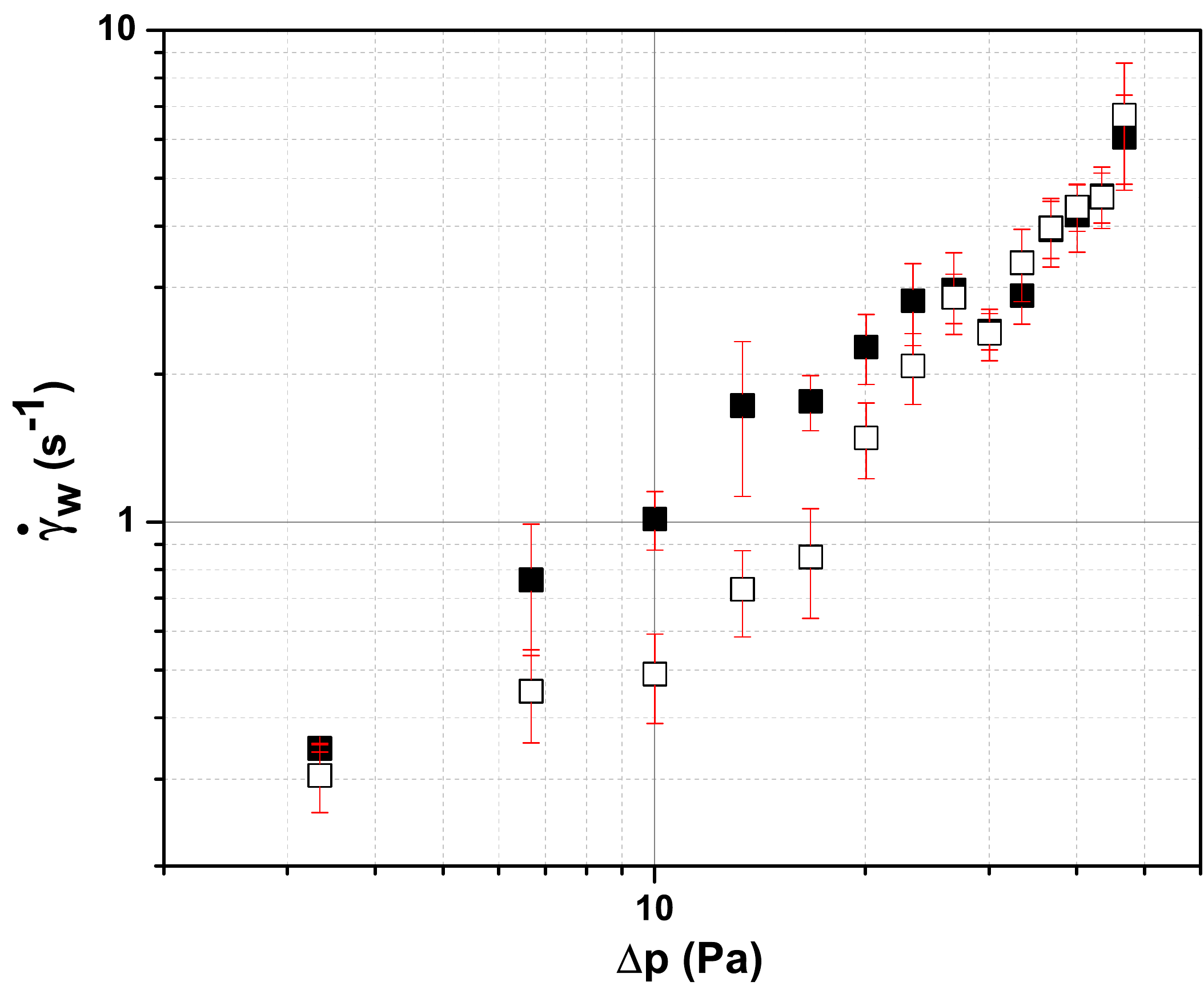}
     \label{fig:wallrate_vs_dp_st}
     }
\subfigure[]{
                 \includegraphics [width=5.7cm] {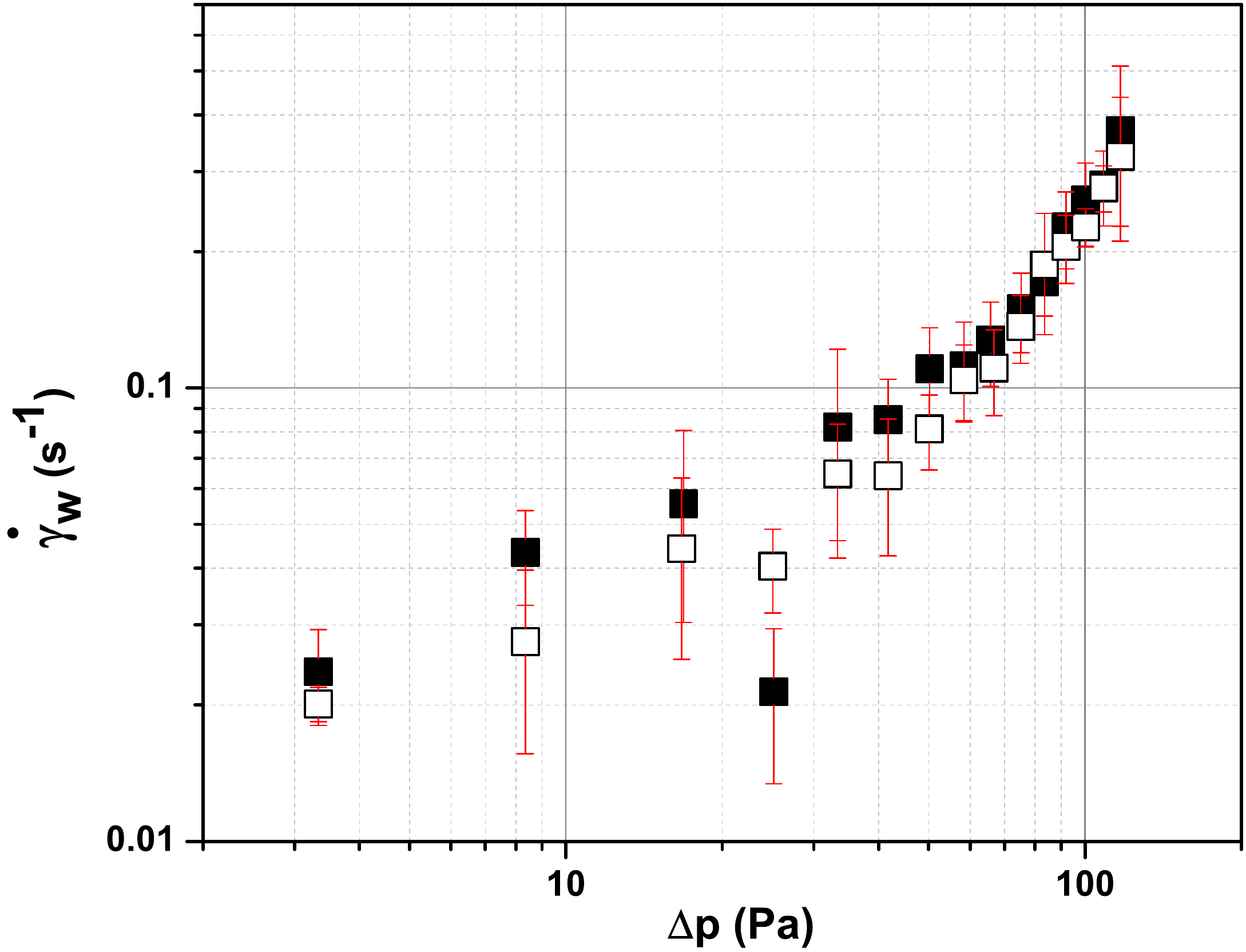}
    \label{fig:wallrate_vs_dp_ys}}
\caption[Dependence of the wall shear rate on the  pressure drop.]{Dependence of the wall shear rate $\dot \gamma_w$ on the driving pressure drop $\Delta p$ for two volume fractions of the suspension: \subref{fig:wallrate_vs_dp_st}  $\Phi_v = 0.12$.  \subref{fig:wallrate_vs_dp_ys}  $\Phi_v = 0.57$. In both panels the full/empty symbols refer to the increasing/decreasing branch of the controlled pressure ramp. The definition of the error bars is given in the text. }
\label{fig:wallrate_vs_dp}
\end{figure*}

The slip velocity may be obtained by extrapolating the time averaged velocity profiles at the positions of the channel's walls,  $y = \pm W/2$. An estimate of the error may be obtained by extrapolating the root mean square deviation (rms) of the velocity measurements (see the error bars in Figs. \ref{fig:profilN}, \ref{fig:profilST}, \ref{fig:profilYS}) at the level of the channel's walls as well. The wall shear stress is often calculated using the pressure drop, the hydraulic diameter and the length of the channel as $\tau_w = \frac{D_h \Delta p}{4 L}$.  In the presence of wall slip, however, this procedure may lead to an underestimation of the true stress at the wall.  Indeed, corresponding to the largest pressure drop explored ($\Delta p = 116.8~Pa$) with the highly concentrated solution with the volume fraction $\Phi_v=0.57$, this estimate would give $\tau_w = 0.071 Pa$ which, according to the rheological measurements presented in Fig. \ref{sample_viscosities}, is roughly an order of magnitude below the yield point. This conclusion is at odds with the measurements of the time averaged profile of the axial velocity presented in Fig. \ref{fig:profilYS} (the squares), the direct visualisation of the flow (data not shown here) and the measurements of the mean flow speed presented in Fig. \ref{fig:averagevelocity} which indicate that, corresponding to this pressure drop, the suspension is partially yielded. 

\begin{figure*}[ht]
\centering
\subfigure[]{
     \includegraphics [width=6cm] {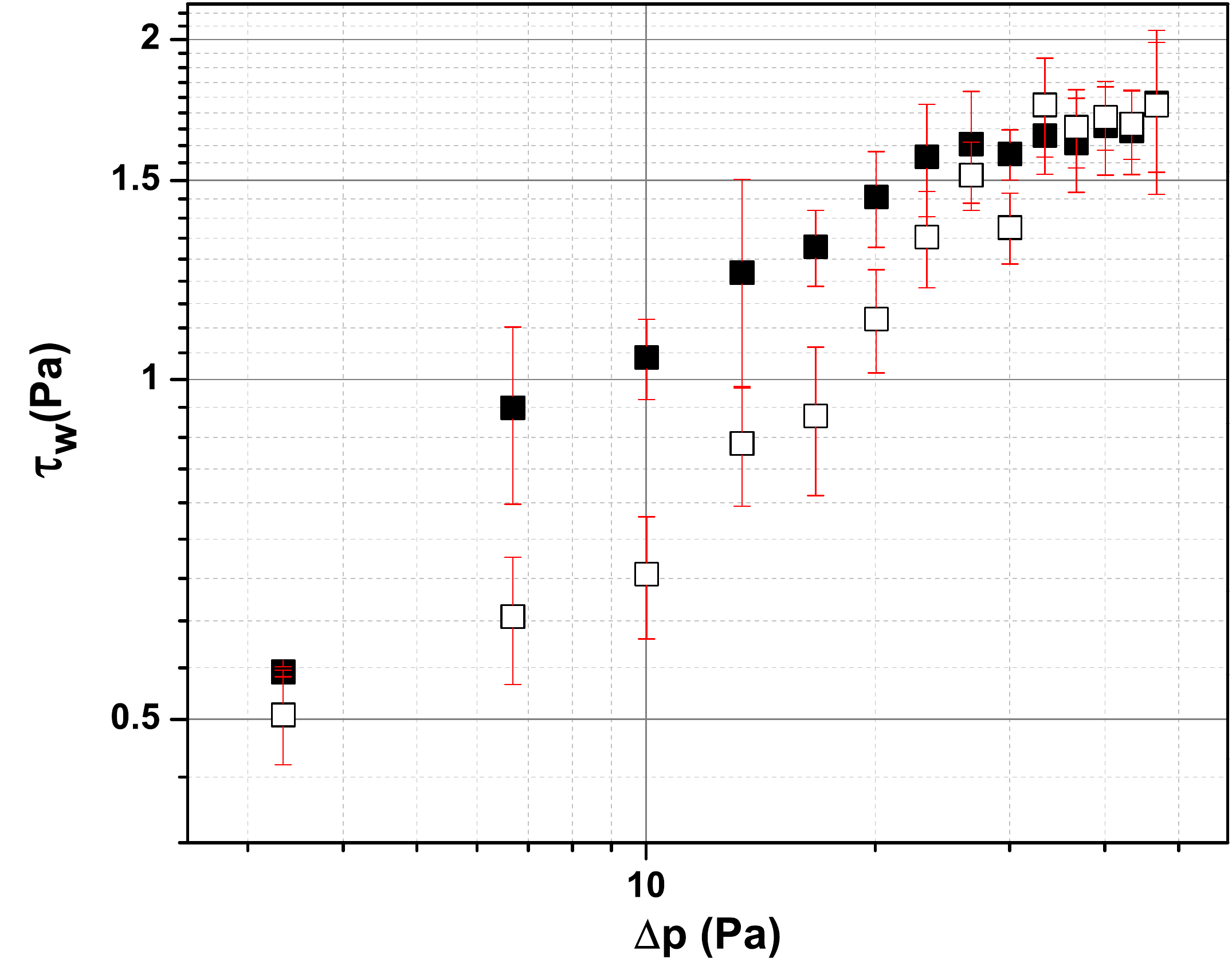}
     \label{fig:wallstress_vs_dp_st}
     }
\subfigure[]{
                 \includegraphics [width=5.8cm] {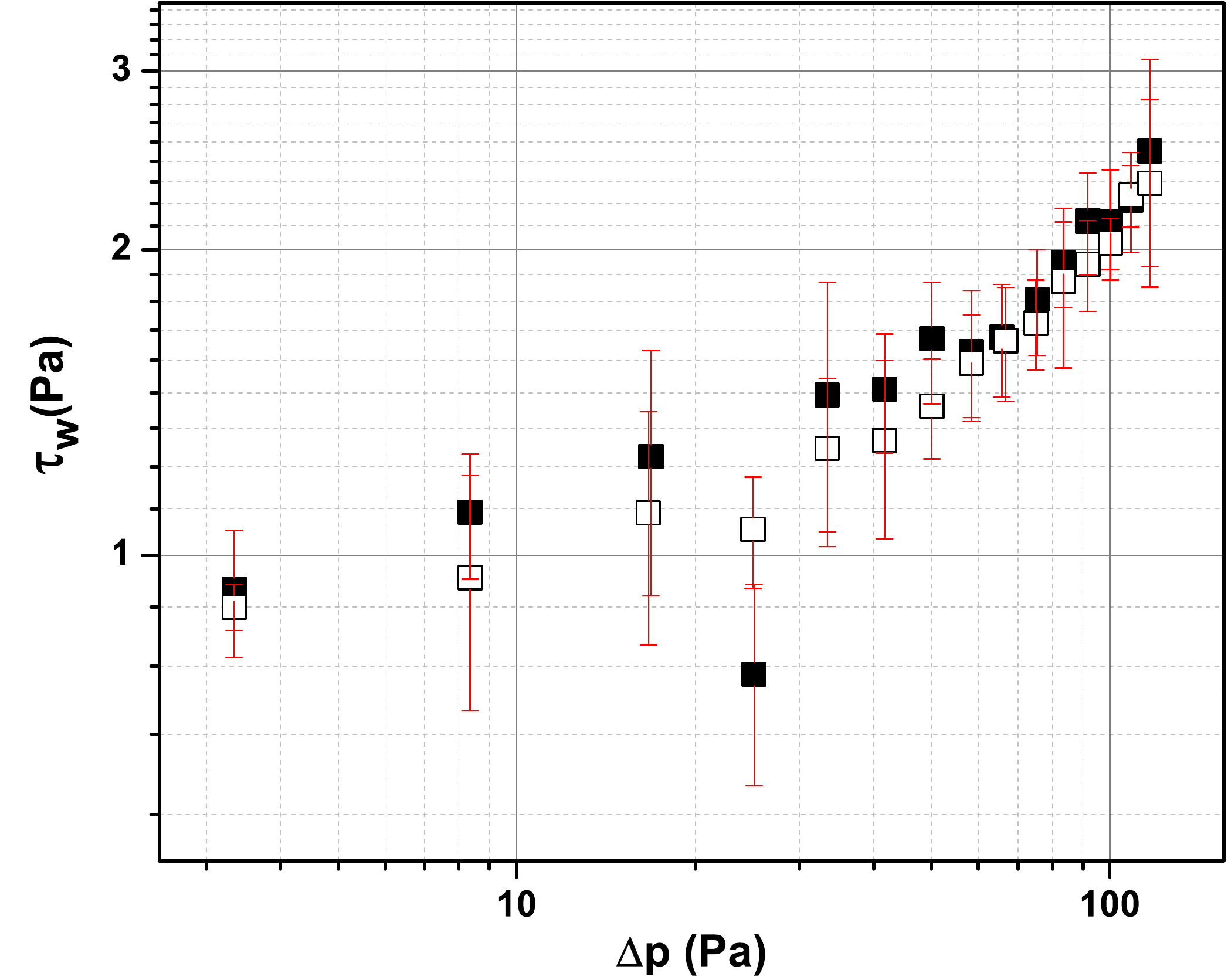}
    \label{fig:wallstress_vs_dp_ys}}
\caption[Dependence of the wall shear stress on the  pressure drop.]{Dependence of the wall shear rate $\dot \gamma_w$ on the driving pressure drop $\Delta p$ for two volume fractions of the suspension: \subref{fig:wallrate_vs_dp_st}  $\Phi_v = 0.12$.  \subref{fig:wallrate_vs_dp_ys}  $\Phi_v = 0.57$. In both panels the full/empty symbols refer to the increasing/decreasing branch of the controlled pressure ramp. The definition of the error bars is given in the text. }
\label{fig:wallstress_vs_dp}
\end{figure*}

To avoid a systematic underestimation of the wall stress we calculate it using the time averaged velocity profiles and the rheological measurements.  
The wall shear rate $\dot \gamma_w$ is obtained by fitting linearly the velocity profiles near the wall, Fig. \ref{fig:wallrate_vs_dp}. As compared to the measurements performed with the shear thinning suspension illustrated in Fig. \ref{fig:wallrate_vs_dp_st} which exhibit a clearly irreversible behaviour upon increasing/decreasing driving pressures, the measurements performed with the yield stress suspension are, within the level of instrumental accuracy indicated by the error bars, reversible Fig. \ref{fig:wallrate_vs_dp_ys}. This is a first indicator of different wall slip mechanisms in the shear thinning and the yield stress cases.

Using the rheological measurements presented in Fig. \ref{sample_viscosities} the wall shear stress may be calculated as $\tau_w = \eta \left ( \dot \gamma_w \right ) \cdot \dot \gamma_w$. The error in the assessment of the wall shear stress may be calculated as  $\delta  \tau_w  =\dot \gamma_w \cdot \delta \eta+ \eta \cdot \delta  \dot \gamma_w$ where $\delta \eta$ is the instrumental error of the viscosity measurement which never exceeded $0.5 \%$ of the measured value and  $\delta  \dot \gamma_w$ is the error of the calculation of the wall shear rate defined by the error of the linear fit of the velocity profile near the wall. The largest contribution to the error in the calculation of the wall stress comes from the error in the calculation of the wall shear rate, particularly at low driving pressure drops.

 The dependence of the calculated wall stress on the pressure drop is presented in Fig. \ref{fig:wallstress_vs_dp}. For the shear thinning suspension (Fig. \ref{fig:wallstress_vs_dp_st}) a clear hysteresis is observed upon increasing/decreasing pressure drops (the full/empty symbols). Due to the rather large instrumental errors, this effect is less obvious for the yield stress suspension (note the overlap of the error bars associated to the full/empty symbols) illustrated in Fig. \ref{fig:wallstress_vs_dp_ys}). It is interesting to compare the measurements of the wall stress performed with the shear thinning and the yield stress suspensions. For the shear thinning case a monotonic increase of the wall stress with the pressure drop is observed up to $\Delta p \approx 25$ followed by a plateau, Fig. \ref{fig:wallstress_vs_dp_st}.  To understand the underlying physics behind the wall shear stress plateau, we first note that slope of the dependence $\tau_w=\tau_w(\Delta p)$  on the pressure drop may be written as $\frac{d \tau_w}{d \Delta p}= \eta_w \frac{d \dot {\gamma_w}}{d \Delta p} + \dot{\gamma_w}\frac{d \eta_w}{d \Delta p}$ where $\eta_w = \eta \vert _{\dot \gamma = \dot \gamma_w}$ is the viscosity of the suspension in the vicinity of the wall which may be readily obtained from the bulk rheological measurements presented in Fig. \ref{sample_viscosities}. In this range of wall shear stresses ($\tau \approx1.5~Pa$), the viscosity exhibits a plateau (see the triangles in Fig. \ref{sample_viscosities}), $\frac{d \eta_w}{d \Delta p} \approx 0$ which leads to   $\frac{d \tau_w}{d \Delta p} \approx \eta_w \frac{d \dot \gamma_w}{d \Delta p}$. The slope of the dependence of the wall shear rate on the driving pressure may be estimated from Fig. \ref{fig:wallrate_vs_dp_st} as $\frac{d \dot \gamma_w}{ d \Delta p} \approx 1.61 s^{-1}Pa^{-1}$ and the corresponding viscosity is $\eta_w \approx 10^{-2} Pas$. Finally, one can estimate $\frac{d \tau_w}{d \Delta p} \approx 0.0161 $. Within the instrumental error of the measurements presented in Fig. \ref{fig:wallstress_vs_dp_st} (note the magnitude of the error bars) this slope is too small to be measured which ultimately explains the observation of wall stress the plateau.
 Like for the wall shear rate data, a clear hysteresis of the wall shear stress is observed upon increasing/decreasing pressure drops.

 In the yield stress case a monotonic increase of the wall stress is observed only beyond a critical value of the pressure drop $\Delta p_0 \approx 30~Pa$. 

\begin{figure*}[ht]
\centering
\subfigure[]{
     \includegraphics [width=6.1cm] {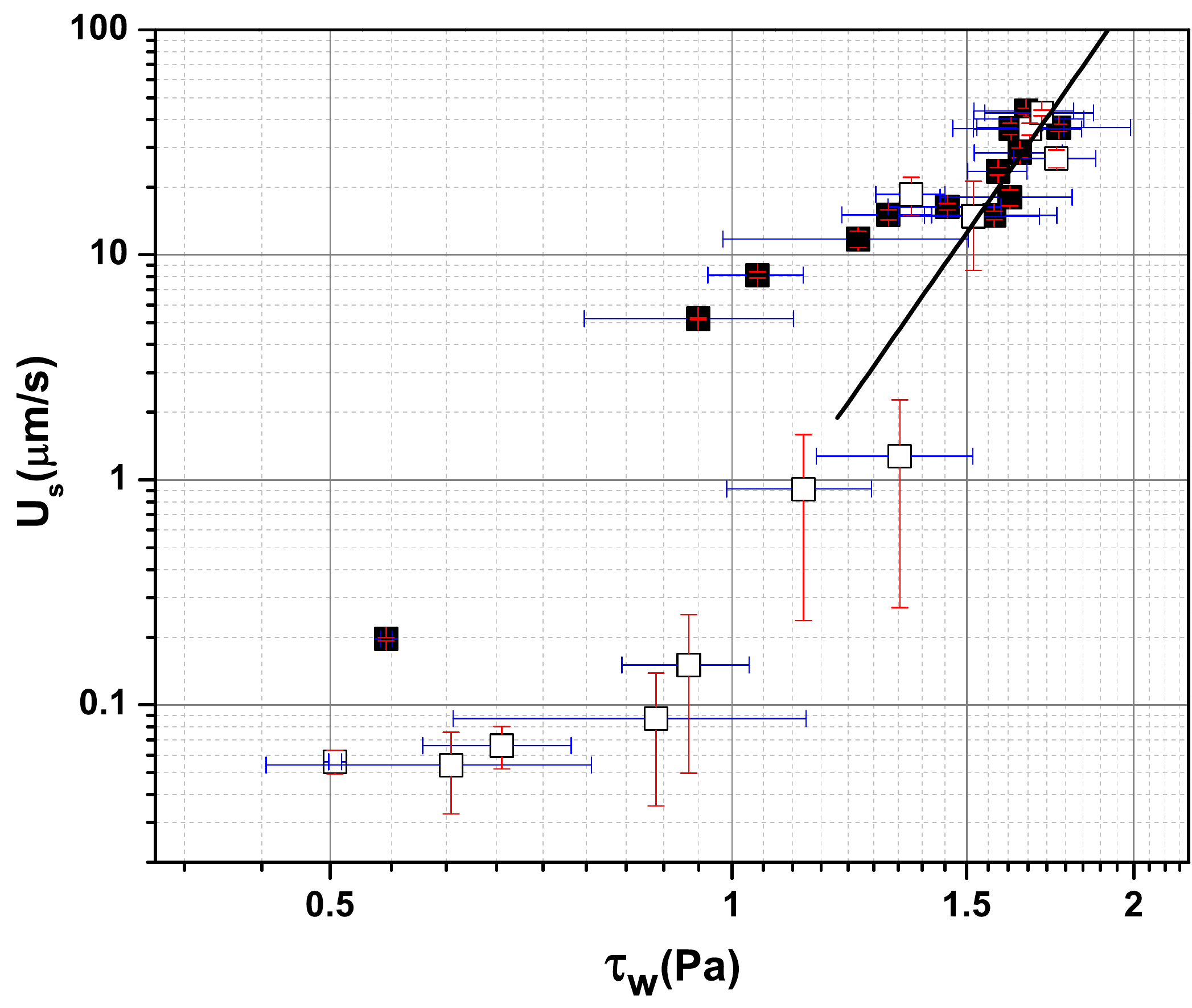}
     \label{fig:slip_vs_stress_st}
}
\subfigure[]{
                 \includegraphics [width=6cm] {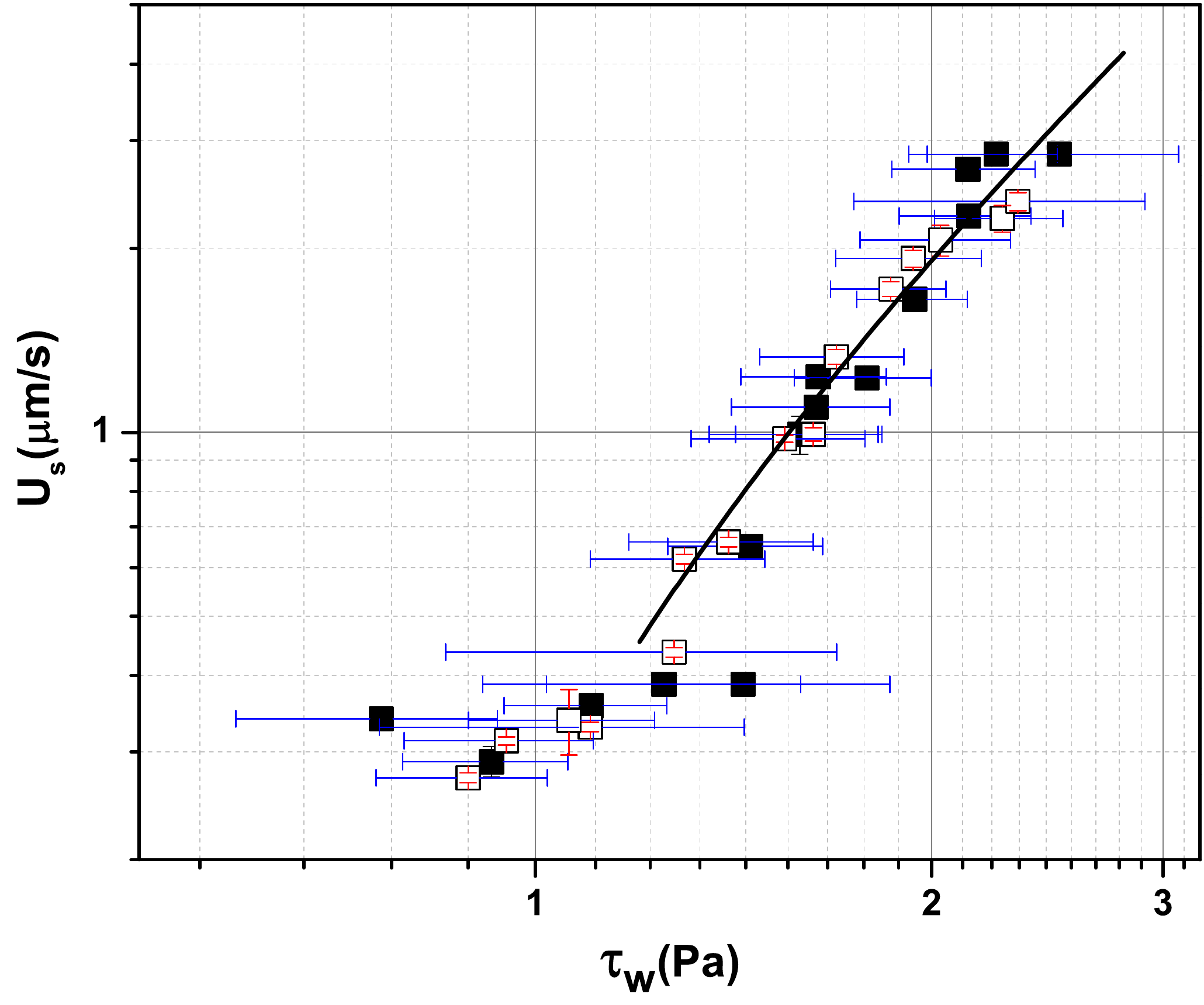}
    \label{fig:slip_vs_stress_ys}}
    \caption[Dependence of the slip velocity on the wall shear stress.]{Dependence of the slip velocity $U_s$ on the wall stress $\tau_w$ for two volume fractions of the suspension: \subref{fig:slip_vs_stress_st}  $\Phi_v = 0.12$.  \subref{fig:slip_vs_stress_ys}  $\Phi_v = 0.57$. The full line  is a power law fit $U_s \propto \tau_w^{\alpha}$ in panel \subref{fig:slip_vs_stress_st} and a nonlinear fit $U_s \propto (\tau_w - \tau_0) ^{\alpha}$ in panel \subref{fig:slip_vs_stress_ys}.  In both panels the full/empty symbols refer to the increasing/decreasing branch of the controlled pressure ramp. The definition of the error bars is given in the text.}
\label{fig:slip_vs_stress}
\end{figure*}

The dependence of the slip velocity on the calculated wall shear stress is presented in Fig. \ref{fig:slip_vs_stress}. For the shear thinning case illustrated in Fig. \ref{fig:slip_vs_stress_st} a large hysteresis is observed in a range of wall stresses $\tau_w \leq 1.5~Pa$. It is interesting to note that beyond this value of stress the bulk rheological measurements presented in Fig. \ref{sample_viscosities} are reversible upon increasing/decreasing stresses (the full/empty triangles). This indicates that in the case of the shear thinning suspension the irreversibility of the slip behaviour is related to the irreversibility of the bulk rheological properties of the suspension. In the reversible range of stresses the slip velocity scales as $U_s\propto \tau_w^{\alpha}$ with $\alpha = 8.5 \pm 0.25$. This unexpectedly steep scaling may be explained by the observation of a plateau of the wall shear stress in this range of driving pressures, the inset in Fig. \ref{fig:wallrate_vs_dp_st}.

A substantially different slip behaviour is observed for the yield stress suspension, Fig. \ref{fig:slip_vs_stress_ys}. First and unlike in the shear thinning case, the wall slip data is practically reversible upon increasing/decreasing pressure drop over the entire range of pressures. At a first glance this observation seems to be at odds with the strong irreversibility of the bulk rheological properties illustrated in Fig. \ref{sample_viscosities} (the full/empty circles). In an attempt to understand this apparent discrepancy we refer to the phenomenological picture of wall slip in concentrated suspensions which exhibit yield stress proposed by Kalyon, \cite{kaylonslip}. According to Kalyon, in the case of a concentrated suspension of rigid particles, the particles can not occupy efficiently the volume elements adjacent to the wall of the channel and thus a particle free layer is formed near the wall. In the case of a concentrated \textit{Chlorella} suspension this layer consists of the culture medium which may be assumed Newtonian. The reversibility of the slip velocity upon increasing/decreasing the wall stress indicates that in the case of a concentrated suspension the slip behaviour is controlled by the Newtonian lubricating layer formed at the wall and not by the bulk dynamics. This conclusion seems to be consistent with the invariance of the plug size $W_p$ with the driving pressure drop observed in Fig. \ref{fig:plugsize} in the range $\Delta p > 70~Pa$. A second interesting observation is that an increase of the slip velocity with the wall stress is observed only beyond a critical value of the wall stress $\tau_w^0$. This prompts us to fit the slip velocity data by $U_s \propto \left ( \tau_w - \tau_w^0\right )^{\alpha}$ which gives $\tau_w^0 = 0.56 \pm 0.045$ and $\alpha = 1.77 \pm 0.3$. The value of the critical stress  $\tau_w^0$ obtained from the nonlinear fit practically coincides with the yield stress of the suspension obtained via a fit of the rheological measurements presented in Fig. \ref{sample_viscosities} by the Herschel-Bulkley model, $\tau_y \approx 0.47~Pa$. 

To sum up, based on the results presented in Figs.  \ref{fig:wallrate_vs_dp_ys}, \ref{fig:slip_vs_stress_ys} and the discussion above, the following phenomenological picture for the flow of the yield stress suspension may be suggested. Corresponding to low pressure drops at which the wall stress does not exceed the bulk yield stress the flow consists of a fully unyielded plug that "slides" on a thin lubricating layer formed by the culture medium along the channel walls. Within this regime, the slip velocity is insensitive to an increase of the wall shear stress. Corresponding to pressure drops sufficiently large so the wall stress exceeds the bulk yield stress a partial yielding of the rigid plug is observed which translates into a monotonic increase of the slip velocity with wall stress. 

\section {Conclusions} \label{sec:conclusions}

A systematic experimental investigation of the microscopic flows of suspensions of \textit{Chlorella} micro-alga is presented. Depending on the volume fraction of the suspension, three flow regimes have been observed. Within a dilute volume fraction regime ($\Phi_v<0.12$) a Newtonian flow behaviour is observed. The transversal velocity profiles can be accurately fitted by the analytical solution of a Poiseuille flow in a rectangular channel  Fig. \ref{fig:profilN} and the flow states are reversible upon increasing/decreasing driving pressures, Figs. \ref{fig:averagevelocity}, \ref{fig:flowresistance} (the full/empty squares). Moreover, no significant slip was observed near the walls of the micro-channel. This flow behaviour is fully consistent with the rheological measurements we have recently reported in Ref. \cite{rheochlorella} within this range of volume fractions. 

As the volume fraction is increased corresponding to the ST rheological regime, significant changes are observed in the flow behaviour. For the largest driving pressures investigated the hydrodynamic reversibility observed in the Newtonian regime is preserved and the transversal velocity profiles can be well described by the Ostwald de Waele law, the full/empty squares in Fig. \ref{fig:profilST}. In a range of low driving pressures, however, the hydrodynamic reversibility is broken and the velocity profiles are no longer described by the Ostwald de Waele law, the full/empty triangles and circles in Fig. \ref{fig:profilST}. In addition to these new flow features, a clear slip is observed at the walls of the micro-channel and its magnitude gradually increases with the driving pressures. For the highest volume fraction investigated corresponding to the YS rheological regime a plug-like microscopic flow consistent with a yield stress rheological behaviour is observed, Fig. \ref{fig:profilYS}.  As for the ST rheological regime, the flow is reversible only for large values of the driving pressures and wall slip is observed in the entire range of pressure drops. The plug flow behaviour may be explained by the formation of a large (comparable in size to the width of the channel) aggregate of small cells, Fig. \ref{fig:plugfield}. The size of the plug decreases with the pressure drop much slower than $1/\Delta p$ reaching a plateau corresponding to $\Delta p \approx 70 Pa$, Fig. \ref{fig:plugsize}. Corresponding to the ST and YS rheological regimes, the microscopic flows are characterised by strongly nonlinear and irreversible flow curves Fig. \ref{fig:averagevelocity} and a large flow resistance, Fig. \ref{fig:flowresistance}. This microscopic flow  behaviour corroborates well with the bulk rheological behaviour probed via increasing/decreasing controlled stress ramps, Fig. \ref{sample_viscosities}. Measurements of the velocity gradients and stresses near the channel wall revealed a highly nontrivial wall slip behaviour corresponding to the ST and YS rheological regimes. In the ST regime the irreversibility of the flow curves visible in Fig.  \ref{fig:averagevelocity} (the up/down triangles) is fully inherited by the walls slip dynamics, Figs. \ref{fig:wallrate_vs_dp_st}, \ref{fig:slip_vs_stress_st}. Bearing in mind that the hydrodynamic irreversibility in this case primarily originates from the irreversibility of the bulk rheological properties upon increasing/decreasing forcing one can conclude at this point that the wall slip dynamics is driven by the bulk flow behaviour. Within the YS rheological regime the wall slip dynamics is substantially different, however. Whereas the bulk rheological properties and the flow curves are irreversible upon increasing/decreasing stresses, the wall slip dynamics is reversible, Fig. \ref{fig:wallrate_vs_dp_ys}, \ref{fig:slip_vs_stress_ys}. A possible explanation for this rather unexpected result may be given in terms of the formation of a cell-free Newtonian fluid layer near the channel walls which controls the slip behaviour. In this case, the bulk of the microscopic flow occupied by a large aggregate of cells plays no significant role in the wall slip dynamics.  Another important result concerning the slip dynamics in the YS regime is that the onset of the slip behaviour practically coincides with the apparent yield stress measured by bulk rheometry, Fig. \ref{fig:slip_vs_stress_ys}. This result indicates that the yielding transition is directly coupled to the wall slip behaviour.

In closing, our findings bring several interesting insights that might be worth being considered by future studies. We have shown that, in spite of an apparently simple cellular structure (particularly the lack of flagella and rigidity of the cellular membrane), the addition of \textit{Chlorella} micro-alga cells to a Newtonian fluid may lead to a rather complex microscopic flow behaviour if the cell volume fraction is sufficiently large, $\Phi_v > 0.1$. We believe that, in the context of the hydrodynamic optimisation of intensified PBR's which is the crucial step for the efficient biofuel production at an industrial scale, this conclusion can not be ignored and there are few directions worth being pursued. First, a better understanding of the coupling between macroscopic rheological properties of the suspensions and microscopic scale interactions/organisations is needed. Second and based on this macroscopic rheological models able to describe the thixotropic like behaviour observed at low applied stresses and/or the shear thinning and yield stress behaviour observed at larger stresses need to be formulated in direct connection with the evolution of the microstructure. Realistic numerical simulations of flows in intensified PBR's and their role in optimising the radiative properties for various lightning conditions (which should ultimately result in  maximising the production of biomass) should be implemented rather than using the commercial CFD tools available on the market.  At a more fundamental level, our systematic experimental observations of hydrodynamic irreversibility upon increasing/decreasing external forcing bear a certain degree of universality and might turn useful in understanding the behaviour of other complex fluids subjected to a varying external stress.   

\section{Acknowledgments}

This work was supported by the PERLE 2 (P\^{o}le d'Excellance pour la Recherche Lig\`{e}rienne en \'{E}nergie) program generously funded by the \textit{Pays de la Loire} district.

\bibliographystyle{model1-num-names}
\hypersetup{citecolor=red}
%\bibliography{ref_micro}

\begin{thebibliography}{36}
\expandafter\ifx\csname natexlab\endcsname\relax\def\natexlab#1{#1}\fi
\providecommand{\bibinfo}[2]{#2}
\ifx\xfnm\relax \def\xfnm[#1]{\unskip,\space#1}\fi
%Type = Article
\bibitem[{Slade and Bauen(2013)}]{Slade201329}
\bibinfo{author}{R.~Slade}, \bibinfo{author}{A.~Bauen},
\newblock \bibinfo{title}{Micro-algae cultivation for biofuels: Cost, energy
  balance, environmental impacts and future prospects},
\newblock \bibinfo{journal}{Biomass and Bioenergy} \bibinfo{volume}{53}
  (\bibinfo{year}{2013}) \bibinfo{pages}{29 -- 38}. \bibinfo{note}{20th
  European Biomass Conference}.
%Type = Article
\bibitem[{Safi et~al.(2014)Safi, Zebib, Merah, Pontalier, and
  Vaca-Garcia}]{chlorella1}
\bibinfo{author}{C.~Safi}, \bibinfo{author}{B.~Zebib},
  \bibinfo{author}{O.~Merah}, \bibinfo{author}{P.-Y. Pontalier},
  \bibinfo{author}{C.~Vaca-Garcia},
\newblock \bibinfo{title}{Morphology, composition, production, processing and
  applications of chlorella vulgaris: A review},
\newblock \bibinfo{journal}{Renewable and Sustainable Energy Reviews}
  \bibinfo{volume}{35} (\bibinfo{year}{2014}) \bibinfo{pages}{265 -- 278}.
%Type = Article
\bibitem[{Al-lwayzy et~al.(2014)Al-lwayzy, Yusaf, and Al-Juboori}]{chlorella2}
\bibinfo{author}{S.~H. Al-lwayzy}, \bibinfo{author}{T.~Yusaf},
  \bibinfo{author}{R.~A. Al-Juboori},
\newblock \bibinfo{title}{Biofuels from the fresh water microalgae chlorella
  vulgaris (fwm-cv) for diesel engines},
\newblock \bibinfo{journal}{Energies} \bibinfo{volume}{7}
  (\bibinfo{year}{2014}) \bibinfo{pages}{1829}.
%Type = Article
\bibitem[{Liang et~al.(2009)Liang, Sarkany, and Cui}]{chlorella3}
\bibinfo{author}{Y.~Liang}, \bibinfo{author}{N.~Sarkany},
  \bibinfo{author}{Y.~Cui},
\newblock \bibinfo{title}{Biomass and lipid productivities of chlorella
  vulgaris under autotrophic, heterotrophic and mixotrophic growth conditions},
\newblock \bibinfo{journal}{Biotechnology Letters} \bibinfo{volume}{31}
  (\bibinfo{year}{2009}) \bibinfo{pages}{1043--1049}.
%Type = Article
\bibitem[{Converti et~al.(2009)Converti, Casazza, Ortiz, Perego, and
  Borghi}]{chlorella4}
\bibinfo{author}{A.~Converti}, \bibinfo{author}{A.~A. Casazza},
  \bibinfo{author}{E.~Y. Ortiz}, \bibinfo{author}{P.~Perego},
  \bibinfo{author}{M.~D. Borghi},
\newblock \bibinfo{title}{Effect of temperature and nitrogen concentration on
  the growth and lipid content of nannochloropsis oculata and chlorella
  vulgaris for biodiesel production},
\newblock \bibinfo{journal}{Chemical Engineering and Processing: Process
  Intensification} \bibinfo{volume}{48} (\bibinfo{year}{2009})
  \bibinfo{pages}{1146 -- 1151}.
%Type = Article
\bibitem[{Kapaun and Reisser(1995)}]{membrane}
\bibinfo{author}{E.~Kapaun}, \bibinfo{author}{W.~Reisser},
\newblock \bibinfo{title}{A chitin-like glycan in the cell wall of a chlorella
  sp. (chlorococcales, chlorophyceae)},
\newblock \bibinfo{journal}{Planta} \bibinfo{volume}{197}
  (\bibinfo{year}{1995}) \bibinfo{pages}{577--582}.
%Type = Article
\bibitem[{Scarsella et~al.(2012)Scarsella, Torzillo, Cicci, Belotti, Filippis,
  and Bravi}]{stresschlorella}
\bibinfo{author}{M.~Scarsella}, \bibinfo{author}{G.~Torzillo},
  \bibinfo{author}{A.~Cicci}, \bibinfo{author}{G.~Belotti},
  \bibinfo{author}{P.~D. Filippis}, \bibinfo{author}{M.~Bravi},
\newblock \bibinfo{title}{Mechanical stress tolerance of two microalgae},
\newblock \bibinfo{journal}{Process Biochemistry} \bibinfo{volume}{47}
  (\bibinfo{year}{2012}) \bibinfo{pages}{1603 -- 1611}.
%Type = Article
\bibitem[{Chisti(2007)}]{Chisti2007294}
\bibinfo{author}{Y.~Chisti},
\newblock \bibinfo{title}{Biodiesel from microalgae},
\newblock \bibinfo{journal}{Biotechnology Advances} \bibinfo{volume}{25}
  (\bibinfo{year}{2007}) \bibinfo{pages}{294 -- 306}.
%Type = Article
\bibitem[{Pulz(2001)}]{Pulz}
\bibinfo{author}{O.~Pulz},
\newblock \bibinfo{title}{Photobioreactors: production systems for phototrophic
  microorganisms},
\newblock \bibinfo{journal}{Applied Microbiology and Biotechnology}
  \bibinfo{volume}{57} (\bibinfo{year}{2001}) \bibinfo{pages}{287--293}.
%Type = Article
\bibitem[{Schenk et~al.(2008)Schenk, Thomas-Hall, Stephens, Marx, Mussgnug,
  Posten, Kruse, and Hankamer}]{Schenk}
\bibinfo{author}{P.~Schenk}, \bibinfo{author}{S.~Thomas-Hall},
  \bibinfo{author}{E.~Stephens}, \bibinfo{author}{U.~Marx},
  \bibinfo{author}{J.~Mussgnug}, \bibinfo{author}{C.~Posten},
  \bibinfo{author}{O.~Kruse}, \bibinfo{author}{B.~Hankamer},
\newblock \bibinfo{title}{Second generation biofuels: High-efficiency
  microalgae for biodiesel production},
\newblock \bibinfo{journal}{BioEnergy Research} \bibinfo{volume}{1}
  (\bibinfo{year}{2008}) \bibinfo{pages}{20--43}.
%Type = Article
\bibitem[{Cornet(2010)}]{cornet1}
\bibinfo{author}{J.-F. Cornet},
\newblock \bibinfo{title}{Calculation of optimal design and ideal
  productivities of volumetrically lightened photobioreactors using the
  constructal approach},
\newblock \bibinfo{journal}{Chemical Engineering Science} \bibinfo{volume}{65}
  (\bibinfo{year}{2010}) \bibinfo{pages}{985 -- 998}.
%Type = Inbook
\bibitem[{Pandey et~al.(2011)Pandey, Larroche, Ricke, and Dussap}]{provost}
\bibinfo{author}{A.~Pandey}, \bibinfo{author}{C.~Larroche},
  \bibinfo{author}{S.~C. Ricke}, \bibinfo{author}{C.~Dussap},
  \bibinfo{title}{Biofuels: Alternative Feedstocks and Conversion Processes},
  \bibinfo{publisher}{Academic Press}.
%Type = Article
\bibitem[{Doucha et~al.(2005)Doucha, Straka, and Livansky}]{doucha}
\bibinfo{author}{J.~Doucha}, \bibinfo{author}{F.~Straka},
  \bibinfo{author}{K.~Livansky},
\newblock \bibinfo{title}{Utilization of flue gas for cultivation of microalgae
  \textit{Chlorella} sp.) in an outdoor open thin-layer photobioreactor},
\newblock \bibinfo{journal}{Journal of Applied Phycology} \bibinfo{volume}{17}
  (\bibinfo{year}{2005}) \bibinfo{pages}{403--412}.
%Type = Article
\bibitem[{Bitog et~al.(2011)Bitog, Lee, Lee, Kim, Hwang, Hong, Seo, Kwon, and
  Mostafa}]{Bitog2011131}
\bibinfo{author}{J.~Bitog}, \bibinfo{author}{I.-B. Lee}, \bibinfo{author}{C.-G.
  Lee}, \bibinfo{author}{K.-S. Kim}, \bibinfo{author}{H.-S. Hwang},
  \bibinfo{author}{S.-W. Hong}, \bibinfo{author}{I.-H. Seo},
  \bibinfo{author}{K.-S. Kwon}, \bibinfo{author}{E.~Mostafa},
\newblock \bibinfo{title}{Application of computational fluid dynamics for
  modeling and designing photobioreactors for microalgae production: A review},
\newblock \bibinfo{journal}{Computers and Electronics in Agriculture}
  \bibinfo{volume}{76} (\bibinfo{year}{2011}) \bibinfo{pages}{131 -- 147}.
%Type = Article
\bibitem[{Pruvost et~al.(2006)Pruvost, Pottier, and Legrand}]{Pruvost20064476}
\bibinfo{author}{J.~Pruvost}, \bibinfo{author}{L.~Pottier},
  \bibinfo{author}{J.~Legrand},
\newblock \bibinfo{title}{Numerical investigation of hydrodynamic and mixing
  conditions in a torus photobioreactor},
\newblock \bibinfo{journal}{Chemical Engineering Science} \bibinfo{volume}{61}
  (\bibinfo{year}{2006}) \bibinfo{pages}{4476 -- 4489}.
%Type = Article
\bibitem[{Doucha and Livansky(2009)}]{cfd1}
\bibinfo{author}{J.~Doucha}, \bibinfo{author}{K.~Livansky},
\newblock \bibinfo{title}{Outdoor open thin-layer microalgal photobioreactor:
  potential productivity},
\newblock \bibinfo{journal}{Journal of Applied Phycology} \bibinfo{volume}{21}
  (\bibinfo{year}{2009}) \bibinfo{pages}{111--117}.
%Type = Article
\bibitem[{Xiong et~al.(2008)Xiong, Li, Xiang, and Wu}]{highdensity}
\bibinfo{author}{W.~Xiong}, \bibinfo{author}{X.~Li},
  \bibinfo{author}{J.~Xiang}, \bibinfo{author}{Q.~Wu},
\newblock \bibinfo{title}{High-density fermentation of microalga chlorella
  protothecoides in bioreactor for microbio-diesel production},
\newblock \bibinfo{journal}{Applied Microbiology and Biotechnology}
  \bibinfo{volume}{78} (\bibinfo{year}{2008}) \bibinfo{pages}{29--36}.
%Type = Article
\bibitem[{Wu and Shi(2008)}]{wu}
\bibinfo{author}{Z.-Y. Wu}, \bibinfo{author}{X.-M. Shi},
\newblock \bibinfo{title}{Rheological properties of \textit{Chlorella
  pyrenoidosa} culture grown heterotrophically in a fermentor},
\newblock \bibinfo{journal}{Journal of Applied Phycology} \bibinfo{volume}{20}
  (\bibinfo{year}{2008}) \bibinfo{pages}{279--282}.
%Type = Article
\bibitem[{Souli\`{e}s et~al.(2013)Souli\`{e}s, Legrand, Provost, Castelain, and
  Burghelea}]{rheochlorella}
\bibinfo{author}{A.~Souli\`{e}s}, \bibinfo{author}{J.~Legrand},
  \bibinfo{author}{J.~Provost}, \bibinfo{author}{C.~Castelain},
  \bibinfo{author}{T.~Burghelea},
\newblock \bibinfo{title}{Rheological properties of suspensions of the green
  micro-alga chlorella vulgaris},
\newblock \bibinfo{journal}{Rheologica Acta}  (\bibinfo{year}{2013}).
%Type = Article
\bibitem[{Scarano and Rhiethmuller(2001)}]{scarano}
\bibinfo{author}{F.~Scarano}, \bibinfo{author}{M.~L. Rhiethmuller},
\newblock \bibinfo{title}{Advances in iterative multigrid piv image
  processing},
\newblock \bibinfo{journal}{Exp. Fluids} \bibinfo{volume}{29}
  (\bibinfo{year}{2001}).
%Type = Book
\bibitem[{Raffel et~al.(2007)Raffel, Willert, Wereley, and Kompenhans}]{piv2}
\bibinfo{author}{M.~Raffel}, \bibinfo{author}{C.~E. Willert},
  \bibinfo{author}{S.~T. Wereley}, \bibinfo{author}{J.~Kompenhans},
  \bibinfo{title}{Particle Image Velocimetry: A Practical Guide (Experimental
  Fluid Mechanics)}, \bibinfo{publisher}{Springer; 2nd edition},
  \bibinfo{year}{September 2007}.
%Type = Phdthesis
\bibitem[{Souli\`{e}s(2014)}]{thesisAntoine}
\bibinfo{author}{A.~Souli\`{e}s}, \bibinfo{title}{Contribution ˆ l\'{e}tude
  hydrodynamique et \`{a} la mod\'{e}lisation des photobior\'{e}acteurs ˆ haute
  productivit\'{e} volumique}, Ph.D. thesis, University of Nantes,
  \bibinfo{year}{2014}.
%Type = Book
\bibitem[{Harris(1989)}]{harris}
\bibinfo{author}{E.~Harris}, \bibinfo{title}{The Chlamydomonas Sourcebook : a
  comprehensive guide to biology and laboratory use},
  \bibinfo{publisher}{Academic Press Inc. San Diego.}, \bibinfo{year}{1989}.
%Type = Article
\bibitem[{Quemada(1997)}]{quemada1}
\bibinfo{author}{D.~Quemada},
\newblock \bibinfo{title}{Rheological modelling of complex fluids. i. the
  concept of effective volume fraction revisited},
\newblock \bibinfo{journal}{The European Physical Journal - Applied Physics}
  \bibinfo{volume}{1} (\bibinfo{year}{1997}) \bibinfo{pages}{119--127}.
%Type = Article
\bibitem[{Simha(1952)}]{simha}
\bibinfo{author}{R.~Simha},
\newblock \bibinfo{title}{A treatment of the viscosity of concentrated
  suspensions},
\newblock \bibinfo{journal}{Journal of Applied Physics} \bibinfo{volume}{23}
  (\bibinfo{year}{1952}) \bibinfo{pages}{1020 -- 1024}.
%Type = Article
\bibitem[{Herschel and Bulkley(1926)}]{hb}
\bibinfo{author}{W.~H. Herschel}, \bibinfo{author}{R.~Bulkley},
\newblock \bibinfo{title}{Konsistenzmessungen von gummi-benzollšsungen},
\newblock \bibinfo{journal}{Kolloid Z.} \bibinfo{volume}{39}
  (\bibinfo{year}{1926}) \bibinfo{pages}{291--300}.
%Type = Article
\bibitem[{Schechter(1961)}]{schechter}
\bibinfo{author}{R.~S. Schechter},
\newblock \bibinfo{title}{On the steady flow of a non-newtonian fluid in
  cylinder ducts},
\newblock \bibinfo{journal}{AIChE Journal} \bibinfo{volume}{7}
  (\bibinfo{year}{1961}) \bibinfo{pages}{445--448}.
%Type = Article
\bibitem[{Kumar et~al.(1981)Kumar, Yadava, and Gaur}]{kumar}
\bibinfo{author}{H.~Kumar}, \bibinfo{author}{P.~Yadava},
  \bibinfo{author}{J.~Gaur},
\newblock \bibinfo{title}{Electrical flocculation of the unicellular green alga
  chlorella vulgaris beijerinck},
\newblock \bibinfo{journal}{Aquatic Botany} \bibinfo{volume}{11}
  (\bibinfo{year}{1981}) \bibinfo{pages}{187 -- 195}.
%Type = Article
\bibitem[{Hadj-Romdhane et~al.(2013)Hadj-Romdhane, Zheng, Jaouen, Pruvost,
  Grizeau, Croué, and Bourseau}]{Hadji}
\bibinfo{author}{F.~Hadj-Romdhane}, \bibinfo{author}{X.~Zheng},
  \bibinfo{author}{P.~Jaouen}, \bibinfo{author}{J.~Pruvost},
  \bibinfo{author}{D.~Grizeau}, \bibinfo{author}{J.~Croué},
  \bibinfo{author}{P.~Bourseau},
\newblock \bibinfo{title}{The culture of chlorella vulgaris in a recycled
  supernatant: Effects on biomass production and medium quality},
\newblock \bibinfo{journal}{Bioresource Technology} \bibinfo{volume}{132}
  (\bibinfo{year}{2013}) \bibinfo{pages}{285 -- 292}.
%Type = Book
\bibitem[{Coussot(2005)}]{coussotbook}
\bibinfo{author}{P.~Coussot}, \bibinfo{title}{Rheometry of pastes, suspensions
  and granular materials}, \bibinfo{publisher}{John Willey \& Sons},
  \bibinfo{year}{2005}.
%Type = Article
\bibitem[{PŽrez-Gonz‡lez et~al.(2012)PŽrez-Gonz‡lez, L—pez-Dur‡n,
  Mar'n-Santib‡–ez, and Rodr'guez-Gonz‡lez}]{pipesteady}
\bibinfo{author}{J.~PŽrez-Gonz‡lez}, \bibinfo{author}{J.~L—pez-Dur‡n},
  \bibinfo{author}{B.~Mar'n-Santib‡–ez},
  \bibinfo{author}{F.~Rodr'guez-Gonz‡lez},
\newblock \bibinfo{title}{Rheo-piv of a yield-stress fluid in a capillary with
  slip at the wall},
\newblock \bibinfo{journal}{Rheologica Acta} \bibinfo{volume}{51}
  (\bibinfo{year}{2012}) \bibinfo{pages}{937--946}.
%Type = Article
\bibitem[{Poumaere et~al.(2014)Poumaere, Moyers-Gonz‡lez, Castelain, and
  Burghelea}]{Poumaere201428}
\bibinfo{author}{A.~Poumaere}, \bibinfo{author}{M.~Moyers-Gonz‡lez},
  \bibinfo{author}{C.~Castelain}, \bibinfo{author}{T.~Burghelea},
\newblock \bibinfo{title}{Unsteady laminar flows of a carbopol gel in the
  presence of wall slip},
\newblock \bibinfo{journal}{Journal of Non-Newtonian Fluid Mechanics}
  \bibinfo{volume}{205} (\bibinfo{year}{2014}) \bibinfo{pages}{28 -- 40}.
%Type = Article
\bibitem[{Sainudiin et~al.(2015)Sainudiin, Moyers-Gonzalez, and
  Burghelea}]{gibbs}
\bibinfo{author}{R.~Sainudiin}, \bibinfo{author}{M.~Moyers-Gonzalez},
  \bibinfo{author}{T.~Burghelea},
\newblock \bibinfo{title}{A microscopic gibbs field model for the macroscopic
  yielding behaviour of a viscoplastic fluid},
\newblock \bibinfo{journal}{Soft Matter} \bibinfo{volume}{11}
  (\bibinfo{year}{2015}) \bibinfo{pages}{5531--5545}.
%Type = Article
\bibitem[{Sainudiin et~al.(2016)Sainudiin, Moyers-Gonzalez, and
  Burghelea}]{gibbs1}
\bibinfo{author}{R.~Sainudiin}, \bibinfo{author}{M.~Moyers-Gonzalez},
  \bibinfo{author}{T.~Burghelea},
\newblock \bibinfo{title}{A nonlinear dynamical system approach for the
  yielding behaviour of a viscoplastic fluid},
\newblock \bibinfo{journal}{http://arxiv.org/pdf/1603.04636}
  (\bibinfo{year}{2016}).
%Type = Article
\bibitem[{Putz and Burghelea(2009)}]{solidfluid}
\bibinfo{author}{A.~M.~V. Putz}, \bibinfo{author}{T.~I. Burghelea},
\newblock \bibinfo{title}{The solid-fluid transition in a yield stress shear
  thinning physical gel},
\newblock \bibinfo{journal}{Rheologica Acta} \bibinfo{volume}{48}
  (\bibinfo{year}{2009}) \bibinfo{pages}{673--689}.
%Type = Article
\bibitem[{Kalyon(2005)}]{kaylonslip}
\bibinfo{author}{M.~Kalyon, Dilhan},
\newblock \bibinfo{title}{Apparent slip and viscoplasticity of concentrated
  suspensions},
\newblock \bibinfo{journal}{Journal of Rheology} \bibinfo{volume}{49}
  (\bibinfo{year}{2005}) \bibinfo{pages}{621--640}.

\end{thebibliography}

\end{document}